\documentclass{iopjournal}

\usepackage[utf8]{inputenc} % allow utf-8 input
\usepackage[T1]{fontenc}    % use 8-bit T1 fonts
\usepackage{url}            % simple URL typesetting
\usepackage{booktabs}       % professional-quality tables
\usepackage{amsfonts}       % blackboard math symbols
\usepackage{nicefrac}       % compact symbols for 1/2, etc.
\usepackage{microtype}      % microtypography
\usepackage{xcolor}         % colors
\usepackage{amsmath}
\usepackage{comment}
\usepackage{graphicx}
\usepackage{subcaption}
\usepackage{float}
\usepackage{natbib}
\usepackage{caption}
\usepackage{booktabs}
\usepackage{pifont}
\usepackage{pifont,xcolor}
\newcommand{\cmark}{{\color{green!60!black}\ding{51}}}  % ✔ in green
\newcommand{\xmark}{{\color{red!70!black}\ding{55}}}    % ✗ in red

\usepackage{scrextend} % for indented blocks - more succinct than itemize
\usepackage{cleveref}
\usepackage{placeins}
\usepackage{colortbl}
\usepackage{multirow}
\usepackage{makecell}
\usepackage{adjustbox}

\usepackage{xfrac}

\usepackage{enumitem}
\setlist{nosep}

\begin{document}

\articletype{Benchmarks} %	 e.g. Paper, Letter, Topical Review...

\title{TCBench: A Benchmark for Tropical Cyclone Track and Intensity Forecasting at the Global Scale}

\author{Milton Gomez$^{1*}$\orcid{0000-0002-0421-3893}, Saranya Ganesh S.$^{1}$, Marie McGraw$^2$, Frederick Iat-Hin Tam$^{1}$,  Ilia Azizi$^{1}$, Samuel Darmon$^{1}$, Monika Feldmann$^{3,5}$, Stella Bourdin$^{4}$,   Louis Poulain-$\,$-Auzéau$^{5}$, Suzana J. Camargo$^{6,7}$, Jonathan Lin$^{8}$, Dan Chavas$^{9}$, Chia-Ying Lee$^{7}$,  Ritwik Gupta$^{10}$, Andrea Jenney$^{11}$, and Tom Beucler$^{1}$ }

\affil{$^1$FGSE/ECCE, University of Lausanne, Lausanne, CH}

\affil{$^2$CIRA, Colorado State University, Fort Collins, CO, USA}

\affil{$^3$University of Bern, Bern, CH}

\affil{$^4$University of Oxford, Oxford, UK}

\affil{$^5$ETH Zürich, Zürich, CH}

\affil{$^6$Climate School, Columbia University, New York, NY, USA}

\affil{$^7$LDEO, Columbia University, New York, NY, USA}

\affil{$^8$Cornell University, Ithaca, NY, USA}

\affil{$^9$Purdue University, West Lafayette, IN, USA}

\affil{$^{10}$University of California, Berkeley, CA, USA}

\affil{$^{11}$Oregon State University, Corvallis, OR, USA}

\affil{$^*$Author to whom any correspondence should be addressed.}

\email{milton.gomez@unil.ch}

% TODO: Update keywords

\begin{abstract}
TCBench is a benchmark for evaluating global, short to medium-range (1-5 days) forecasts of tropical cyclone (TC) track and intensity. To allow a fair and model-agnostic comparison, TCBench builds on the IBTrACS observational dataset and formulates TC forecasting as predicting the time evolution of an existing tropical system conditioned on its initial position and intensity. TCBench includes state-of-the-art physics-based (TIGGE) and Artificial Intelligence Weather Prediction (AIWP) models (AIFS, Pangu-Weather, FourCastNetV2, GenCast, FNV3P2). If not readily available, TC tracks are consistently derived from model outputs using the TempestExtremes library. TCBench provides deterministic and probabilistic storm-following metrics. On 2023 test cases, AIWP models skillfully forecast TC tracks, while realistic intensity forecasts require additional steps such as post-processing or task-specific training. Designed for accessibility, TCBench helps AI practitioners tackle domain-relevant TC challenges and equips tropical meteorologists with data-driven tools and workflows to improve prediction and TC process understanding. By lowering barriers to reproducible, process-aware evaluation of extreme events, TCBench aims to democratize data-driven TC forecasting.
\end{abstract}

\section{Introduction}

\begin{figure}
  \centering
  \includegraphics[width=0.92\textwidth]{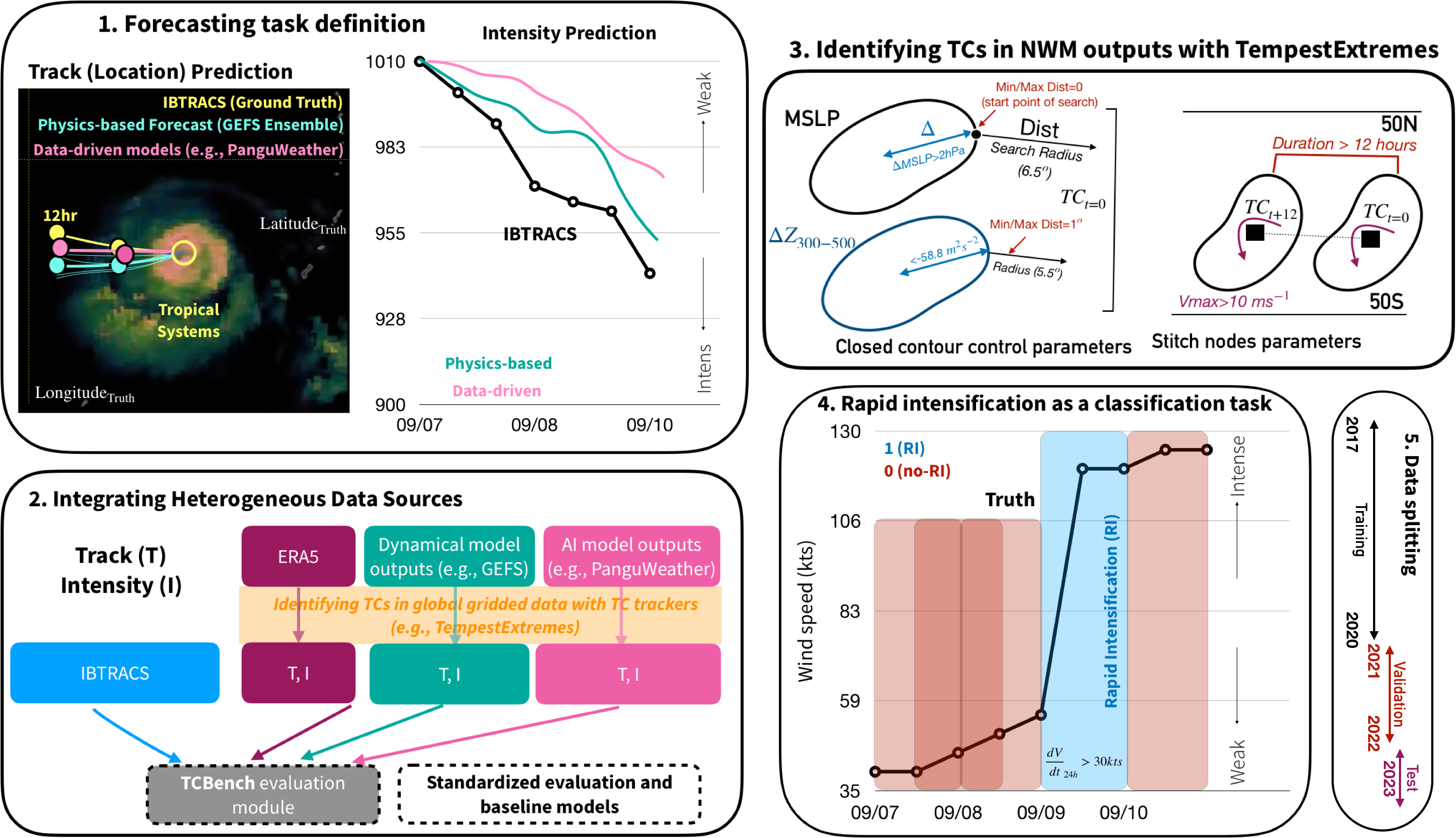}
  \caption{TCBench defines TC forecasting as predicting time-series of track and intensity knowing the system’s initial state. It integrates heterogeneous data sources (observations, reanalysis, physics/data-driven models) into a unified evaluation framework to standardize model assessment.}
  \label{fig:schematic}
\end{figure}

Tropical cyclones (TCs), also called ``hurricanes'' or ``typhoons'' depending on the basin~\citep{Section2_AMS_tropicalcyclone}, are globally devastating weather systems. In the United States alone, TCs caused over \$1.5 trillion in damages and over 7,000 deaths between 1980 and 2024~\citep{NOAAbillion}. With over half of the global population projected to live in the tropics and low-elevation coastal zones by 2050-2060~\citep{gu2021major,neumann2015future}, improving TC forecasting is urgent for risk mitigation and community resilience. 

%% DEFINE TC INTENSITY AND TRACK IN A MORE CONCISE WAY
TC forecasting can be framed as the task of forecasting two continuous time-series: 1) the track, i.e., the time series of storm locations in latitude $\phi \in [-50,50]^\circ$ and longitude $\lambda \in [0,360]^\circ$, and 2) the intensity: commonly characterized by maximum sustained wind speed $V_{\text{max}}$ (m s$^{-1}$) and minimum sea-level pressure $p_{\text{min}}$ (Pa)--noting that both wind and pressure are used operationally.

%\begin{addmargin}[1.5em]{0em}
%\textbf{Track}: the time-series of storm locations in latitude $\phi \in [-50, 50]^\circ$ and longitude $\lambda \in [0, 360]^\circ$; \vspace{\listspacing}\\
%\textbf{Intensity}: commonly characterized by maximum sustained wind speed $V_{\text{max}} \in \mathbb{R}_+$ (m s$^{-1}$) and minimum sea-level pressure $p_{\text{min}} \in \mathbb{R}_+$ (Pa). Both variables are used operationally.   

% \begin{addmargin}[1.5em]{0em}
% Track: the time series of storm locations in latitude $\phi \in [-50,50]^\circ$ and longitude $\lambda \in [0,360]^\circ$.

% \noindent Intensity: commonly characterized by maximum sustained wind speed $V_{\text{max}}$ (m s$^{-1}$) and minimum sea-level pressure $p_{\text{min}}$ (Pa). Both variables are used operationally.
% \end{addmargin}

 %recent work suggests $p_{\text{min}}$ may better correlate with storm damage \citep{Section2_Klotzbach2020}. 

% MOTIVATE DATA-DRIVEN MODELS
Operational TC forecasts traditionally rely on physics-based models that solve coupled partial differential equations to simulate global weather fields, with nested higher-resolution grids in the vicinity of the TC~\citep{HAFS}. However, supercomputing infrastructure is required to run these models at a spatial resolution that marginally resolves TCs~\citep{davis2018resolving} and this limits their accessibility. 
This limitation, combined with the availability of high-quality, open-sourced benchmark datasets~\citep{rasp2024weatherbench}, has led to an increased interest in data-driven weather forecasting~\citep{ebert2023outlook}.

%However, the to the availability of high-quality, open-sourced benchmark datasets \citep{rasp2024weatherbench}, has led to an increased interest in data-driven weather forecasting \citep{ebert2023outlook}. 
Several ``neural weather models''---Artificial Intelligence Weather Prediction (AIWP) models that rely on deep learning algorithms---now outperform state-of-the-art physics models on the 1–10 day forecasting of select global meteorological fields at a $0.25^\circ$ resolution~\citep{Rasp2024livingreview}. Compared to physics-based models, neural weather models---once trained--- are computationally efficient and help democratize forecasting. Therefore, neural weather models can support the generation of large calibrated ensembles~\citep{Brenowitz2025ProbBenchmark}, which is a desirable property in weather forecasting to quantify and reduce uncertainty. Nonetheless, the performance of neural weather models is often summarized in metrics calculated over the global prediction, less focused on process assessment, and divergent approaches in evaluation complicates comparison between models. For example, Pangu-Weather~\citep{Section4_bi2023panguweather} and FourCastNet~\citep{Section4_pathak2022fourcastnet} rely on author-defined TC tracking algorithms, GraphCast~\citep{Section4_lam2023graphcast} uses tracking algorithms optimally selected for each of their model and baseline, and GenCast~\citep{price2025probabilistic} relies on a well-established tracking algorithm~\citep{tempestextremes2021} but is forced to adapt it due to the frequency of their model outputs. Furthermore, other data-driven approaches that focus on basin-specific predictions~\citep[e.g.,][]{huang2023mgtcf, park2023long} provide specialized predictions that may outperform global approaches, given the variance in meteorological and oceanographic characteristics that drive TC behavior in each basin~\citep{singh2025tropical}.

%However, much of the existing evaluation of neural network models focus mostly on weather averaged over the entire globe. There is still a lack of \textbf{unified, process-based assessments} on the performance on local extreme weather like tropical cyclones. 

While TC track prediction from neural weather models now rivals~\citep{demaria2024evaluationtropicalcyclonetrack} and sometimes outperforms physics-based models~\citep{broad2024aiweather}, TC intensity prediction remains a major challenge. %\rev{
Intensity change of TCs is primarily determined by the oceanic energy available for the TC to extract and the vertical structure of temperature in the vicinity of the TC~\citep{emanuel1986air}. It is thus possible to derive the theoretical maximum energy achievable by a TC based on sea level pressure, sea surface temperature, relative humidity, and outflow layer temperature~\citep[MPI--maximum potential intensity; e.g., ][]{emanuel1999thermodynamic}. Furthermore, the main processes that keep TCs from intensifying and reaching their theoretical MPIs include vertical wind shear, which introduces dry air intrusion into the TC~\citep{wong2004tropical,alland2021combined}, and upwelling of cold deep ocean water underneath the TC~\citep{price1981upper}. Though some recent models explicitly predict an ocean variable~\citep[e.g., sea surface temperature][]{price2025probabilistic}, most predict strictly atmospheric data. Given that the state of the ocean can have a strong impact on TC intensity even under adverse atmospheric conditions~\citep{nickerson2025rapid}, this is likely to present an additional challenge for accurately representing intensity. Furthermore, %}
the spatial resolution of physics-based global models is often too coarse to represent anomalies in wind speed, wind extremes, in TCs~\citep{Section2_DeMaria_etal2014,baker2024TCintensification}. Similarly, neural models often forecast low intensity due to biases in their training data~\citep{dulac2024assessing}. As a result, current data-driven intensity forecasts frequently under-perform simple baselines such as persistence and climatology~\citep{demaria2024evaluationtropicalcyclonetrack}.

Accurate modeling of TC intensity is of special importance, given that Rapid Intensification (RI)---defined as a large increase in intensity over a short period of time, typically around the 95\textsuperscript{th} percentile of intensification in a 24h window---is frequently observed in the most destructive TCs %\rev{
\citep{lockwood2024increasing, Knutson2024gfdl} and the frequency of such storms is projected to increase with anthropogenic warming~\citep{li2023recent}.
%}
%\rev{
Furthermore, RI forecasting remains a major challenge because it arises from nonlinear interactions between environmental forcing and inner-core dynamics~\citep{wang2004current}. These multiscale processes involve inner-core features such as eyewall dynamics, vortex Rossby waves, and rainband structures, along with external influences including vertical wind shear and synoptic flow. The balance between these internal and external factors governs how efficiently a storm can intensify, and their nonlinear coupling across scales makes RI particularly difficult to predict. %}
Yet, the predictive performance of neural weather models for RI has to our knowledge rarely been systematically assessed in previous literature. In this work, we treat RI as a binary classification task, which determines whether $V_{\text{max}}$ will increase by at least $ 30~$kt ($\approx15.4~$m s$^{-1} $) over 24 hours~\citep{kaplan2003,Kaplanetal2010}, emphasizing that this roughly corresponds to the 95th percentile of the rate of intensity increase. %\rev{
By formulating RI evaluation in this way, we aim to evaluate how these models could be used as flags in warning systems. This is important given the need for better forecasting of these events to coordinate prompt responses and minimize the impacts associated with these storms~\citep{appendini2024developing}.
%}

%We expect this progress to continue with the planned release of improved meteorological products such as ERA6 \citep{hersbach2024era6}.

% Tom: We need to clearly state that RI is not one of the 4 continuous variables but "included in the prediction of V_max" to avoid confusing reviewers.
%A key sub-task \textbf{Rapid intensification (RI)}, defined as a large increase in intensity over a short period of time. In TCBench, we define RI as an increase in $V_{\text{max}}$ of at least $ 30~$kt ($\approx15.4~$m s$^{-1} $) over 24~hours~\citep{kaplan2003}, roughly corresponding with the 95th percentile of the rate of intensity increase. We note that while we use a threshold applied by the National Hurricane Center in the U.S., other regions and agencies may use other thresholds \citep{Kaplanetal2015}. In TCBench, we frame RI prediction as a \textbf{binary classification task}: determining whether a TC will undergo RI within a given 24-hour window~\citep{Kaplanetal2010}. 

In this work, we present TCBench, a benchmark dataset that bridges state-of-the-art neural and physics-based weather models with TC observational records to accelerate progress in data-driven forecasting of TCs. This enables a transparent, reproducible assessment of how TC prediction can be further improved with machine learning. We demonstrate that neural weather models can skillfully forecast TC intensity up to five days ahead, particularly when combined with observational data and processed through tools provided in TCBench. Furthermore, we show that ensemble neural weather models demonstrate TC track prediction skill that rivals traditional, physics-based ensembles. %, the neural weather model track ensembles rival those of physics-based ensembles in accuracy.

TCBench frames the problem of TC forecasting under the assumption that there is already a tropical system that exists at an initial time of forecast, thereby focusing on predicting how the system will evolve. TCBench thus focuses on the performance of models across all observed storms and not the intersection of storms predicted by each model (i.e., the storms the models are evaluated on are not \textit{homogenized}). We consider this a necessity when comparing a large number of models with different biases (stemming from, e.g., tracking choices), as the size of the evaluation set would otherwise be constrained by the most restrictive model. TCBench provides pre-processing pipelines, evaluation protocols, visualization tools, and baselines from both physics- and neural-based models to benefit the atmospheric science and AI communities. Designed for rapid iteration and anticipating fast innovation in data-driven forecasting, TCBench is extensible to new models, additional predictive targets, and encourages creative use of the data's spatiotemporal structure.

\section{Related Work}
% Do we need this in the paper when we are not using SHIPS at all?
Statistical models skillfully forecast TC intensity up to five days ahead when combined with physics-based global atmospheric models, a strategy known as ``statistical-dynamical'' prediction. A prominent example is the Statistical Hurricane Intensity Prediction Scheme---SHIPS\citep{Section2_DeMaria_Kaplan1994,Kaplan+DeMaria1999, DeMaria_etal_2005_SHIPS}, which uses a multiple regression model with area-averaged climatological, persistence, satellite observations, and environmental predictors from physics-based models to forecast TC intensity. 
%The SHIPS framework spawned additional statistical-dynamical models using similar variables to SHIPS but different statistical methods, such as the Logistic Growth Equation Model (LGEM, \citep{DeMaria2009}), and a version of SHIPS that explicitly includes the effects of land (DSHIPS, \citep{DeMariaetal2005}). 
Statistical-dynamical TC intensity prediction models remain in use operationally due to their skill, particularly at longer forecast lead times~\citep{Cangialosietal2020}. Statistical-dynamical models are also skillful at predicting RI, unlike purely physics-based models~\citep{Torn+DeMaria2021}. Statistical-dynamical methods for RI forecasts, such as the SHIPS RI Index---SHIPS-RII~\citep{Kaplanetal2010,Kaplanetal2015}---use linear discriminant analysis~\citep{Kaplanetal2010,Knaffetal2018}, Bayesian and logistic regression models~\citep{Rozoff+Kossin2011,Knaffetal2018}, binomial logistic regression~\citep{DeMariaetal2021}, as well as consensus models~\citep{Kaplanetal2015}.

In recent years, machine learning models have begun tackling the problem of TC intensity prediction. Machine learning models trained on SHIPS predictors and satellite imagery skillfully predict TC intensity and RI~\citep{Suetal2020,Griffinetal2022}, as well as intensity and track errors~\citep{Barnesetal2023,Fernandezetal2025}. Convolutional neural networks (CNNs) and k-means clustering better distinguish between environments that are favorable and not favorable to RI~\citep{Merceretal2021}. Other approaches, such as decision tree-based models trained on SHIPS predictors~\citep{Shaiba+Hahsler2016}, the random Forest-based RI Scheme (FRIA)~\citep{Slocum2021,Sampsonetal2023}, and the Long Short-Term Memory method~\citep{Yang_etal_2020_RI} have also shown skill in RI prediction.  
The advent of neural weather models for the global atmosphere opens new opportunities for data-driven post-processing, analogous to statistical-dynamical prediction, with the potential to significantly extend the lead time for skillful TC forecasts.

%TCBench stands out by focusing on global TC forecasting and open, real-time–available data, packaged in an accessible format designed for the improvement of data-driven models (Table~\ref{tab:tc_comparison}).
TCBench stands out by targeting the challenging problem of global TC forecasting, providing open, \mbox{real-time–available data} in an accessible format designed to accelerate improvement of data-driven models (\Cref{tab:tc_comparison}). As key contributions from TCBench to the community, we include routines to deterministically and probabilistically evaluate tracks predicted by models, neural weather model output data associated with our selected test year, and the necessary inputs for the training and validating post-processing baselines---facilitating evaluation by community members of their own models and approaches with respect to the provided baselines. Furthermore, our forecasting setup provides a straightforward comparison between TC track prediction models to our best observational record, and does not limit comparisons to storms predicted by all models. Finally, we provide scores for reference models (``baselines'') and thereby streamline comparisons between current and future models. 

% \begin{table}[ht]
% \centering
% \begin{tabular}{lcccccc}
% \toprule
% \textbf{Dataset} &
% \begin{tabular}[c]{@{}c@{}}Forecast\\eval.\end{tabular} &
% \begin{tabular}[c]{@{}c@{}}Global\\coverage\end{tabular} &
% \begin{tabular}[c]{@{}c@{}}Open\\source\end{tabular} &
% \begin{tabular}[c]{@{}c@{}}TC-\\specific\end{tabular} &
% \begin{tabular}[c]{@{}c@{}}Environ.\\predictors\end{tabular} &
% \begin{tabular}[c]{@{}c@{}}AI-ready\\format\end{tabular} \\
% \midrule
% WeatherBench 2 & \cmark & \cmark & \cmark & \xmark & \cmark & \cmark \\
% ChaosBench~\citep{chaosbench}              & \cmark & \cmark & \cmark & \xmark & \cmark & \cmark \\
% SHIPS~\citep{deMaria2022}                        & \xmark & \xmark & \xmark & \cmark & \cmark & \xmark \\
% IBTrACS~\citep{knapp2010international}                    & \xmark & \cmark & \cmark & \cmark & \xmark & \xmark \\
% TC-PRIMED~\citep{TCPRIMED}                 & \xmark & \cmark & \cmark & \cmark & \cmark & \cmark \\
% TropiCycloneNet~\citep{huang2025benchmark}                 & \xmark & \cmark & \cmark & \cmark & \cmark & \cmark \\
% \textbf{TCBench (this work)}                   & \cmark & \cmark & \cmark & \cmark & \cmark & \cmark \\
% \bottomrule
% \end{tabular}
% \caption{Comparison of datasets and benchmarks across key criteria: forecast evaluation support, coverage, openness, TC specificity, environmental predictors, and ML readiness.}
% \label{tab:tc_comparison}
% \end{table}

\begin{table}[ht]
    \centering
    \small
    \setlength{\tabcolsep}{6pt}
    \renewcommand{\arraystretch}{1.15}
    \begin{tabular}{@{}lcccccc@{}}
        \toprule
        \textbf{Dataset} &
        \begin{tabular}[c]{@{}c@{}}Forecast\\evaluation\end{tabular} &
        \begin{tabular}[c]{@{}c@{}}Global\\coverage\end{tabular} &
        \begin{tabular}[c]{@{}c@{}}Open\\source\end{tabular} &
        \begin{tabular}[c]{@{}c@{}}TC-\\specific\end{tabular} &
        \begin{tabular}[c]{@{}c@{}}Environment\\predictors\end{tabular} &
        \begin{tabular}[c]{@{}c@{}}AI-ready\\format\end{tabular} \\
        \midrule
        WeatherBench 2~\citep{rasp2024weatherbench} & \cmark & \cmark & \cmark & \xmark & \cmark & \cmark \\
        ChaosBench~\citep{chaosbench} & \cmark & \cmark & \cmark & \xmark & \cmark & \cmark \\
        SHIPS~\citep{deMaria2022} & \xmark & \xmark & \xmark & \cmark & \cmark & \xmark \\
        IBTrACS~\citep{knapp2010international} (``ground truth'')  & \xmark & \cmark & \cmark & \cmark & \xmark & \xmark \\
        TC-PRIMED~\citep{TCPRIMED} & \xmark & \cmark & \cmark & \cmark & \cmark & \cmark \\
        TropiCycloneNet~\citep{huang2025benchmark}   & \xmark & \cmark & \cmark & \cmark & \cmark & \cmark \\
        Extreme Weather Bench~\citep{EWBench} & \cmark & \cmark & \cmark & \cmark & \xmark & \cmark \\
        CausalLearningAI TCBench~\citep{CausalLearningAI_tcbench}\textsuperscript{\dag} & \xmark & \cmark & \cmark & \cmark & \cmark & \cmark \\
        % WeatherBench 2\textsuperscript{1} & \cmark & \cmark & \cmark & \xmark & \cmark & \cmark \\
        % ChaosBench\textsuperscript{2}              & \cmark & \cmark & \cmark & \xmark & \cmark & \cmark \\
        % SHIPS\textsuperscript{3}                    & \xmark & \xmark & \xmark & \cmark & \cmark & \xmark \\
        % IBTrACS\textsuperscript{4} (``ground truth'')                  & \xmark & \cmark & \cmark & \cmark & \xmark & \xmark \\
        % TC-PRIMED\textsuperscript{5}                & \xmark & \cmark & \cmark & \cmark & \cmark & \cmark \\
        % TropiCycloneNet\textsuperscript{6}          & \xmark & \cmark & \cmark & \cmark & \cmark & \cmark \\
        \textbf{TCBench}                & \cmark & \cmark & \cmark & \cmark & \cmark & \cmark \\
        \bottomrule
    \end{tabular}
    \caption{Comparison of datasets and benchmarks across key criteria: forecast evaluation support, coverage, openness, TC specificity, environmental predictors, and ML readiness.\\
    {\footnotesize \textsuperscript{\dag}project bears no relation to this article}}
    \label{tab:tc_comparison}
    % \vspace{2pt}
    % \par\footnotesize
    % \textsuperscript{1}\,\citep{rasp2024weatherbench};
    % \textsuperscript{2}\,\citep{chaosbench};
    % \textsuperscript{3}\,\citep{deMaria2022};
    % \textsuperscript{4}\,\citep{knapp2010international};
    % \textsuperscript{5}\,\citep{TCPRIMED};
    % \textsuperscript{6}\,\citep{huang2025benchmark}.
\end{table}

\section{TCBench Data Toolbox}\label{sec:DataToolbox}
\begin{figure}[hb!]
\centering
\includegraphics[width=\textwidth]{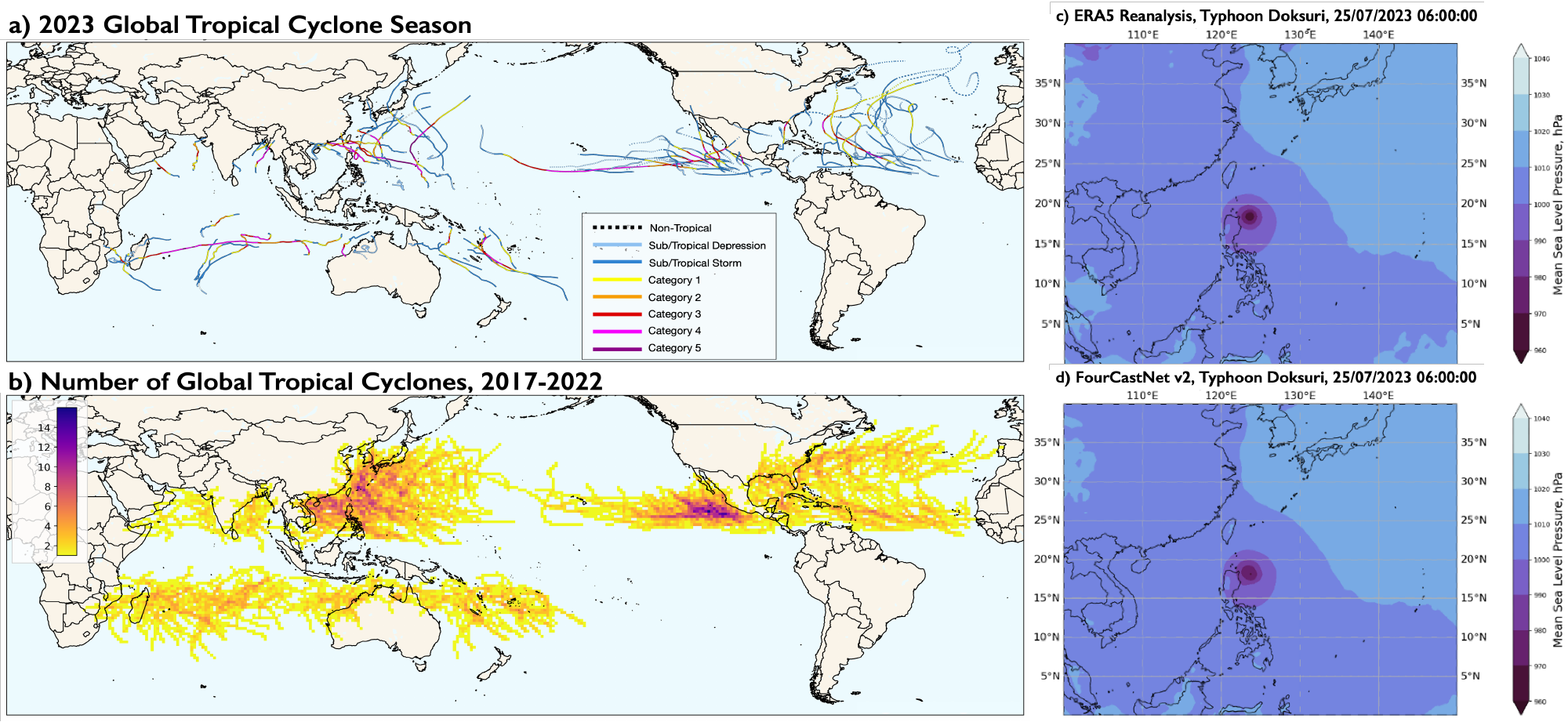}
\caption{ \textbf{(a)} 2023 tropical cyclones in the TCBench test year, from IBTrACS. The lines represent the position of each tropical cyclone over time (``tracks''), with the line color representing the storm's intensity at that position.
\textbf{(b)} IBTrACS estimate of tropical cyclone numbers from 2017-2022 (corresponding to TCBench's training and validation years). Tropical cyclone counts binned into a 1$^\circ$ latitude by 1$^\circ$ longitude grid.
\textbf{(c)} Mean sea level pressure from ERA5 reanalysis during Typhoon Doksuri, 25/07/2023 06:00:00 UTC. Represents the ground truth targeted by neural weather models, as opposed to the best record of TC intensity (IBTrACS)
\textbf{(d)} 6-hour forecast of sea level pressure from the FourCastNetV2 neural weather model during Typhoon Doksuri, valid at 25/07/2023 06:00:00~UTC.
Panels a) and b) were made using the troPYcal software package~\citep{tropycal}.}
\label{fig:datasets}
\end{figure}
%\subsection{TC Tracks in TCBench}
%\subsection{Data Toolbox}
To provide a clear test set and baselines ensuring fair inter-model comparison, TCBench integrates a diverse set of climate data with physics-based and neural weather model forecasts. A summary of all data sources used in TCBench is provided in Appendix~C. These sources were selected based on availability, spatial resolution, historical record length, and their relevance to forecasting or post-processing TC intensity or tracks (\Cref{sec:inclusion_criteria}). \Cref{fig:datasets} provides an example of some of the various data sources included in TCBench, including observed TC track and intensity data (\Cref{fig:datasets}a, \ref{fig:datasets}b), reanalysis (\Cref{fig:datasets}c), and neural weather model forecasts (\Cref{fig:datasets}d). In addition to the data we provide, we set forth guidelines for community members to submit data and/or models for their evaluation and listing on the leaderboard. 
%\subsection{Observations and Reanalysis}

\subsection{Data Provided with TCBench}
The International Best Track Archive for Climate Stewardship (IBTrACS) serves as the ``ground truth'' for model evaluation. IBTrACS is the most complete observational archive of global TCs that is currently available~\citep{knapp2010international,gahtan2024ibtracs}. IBTrACS provides global coverage of TC tracks, including various parameters, such as location, intensity, and size. For consistency, we retain only the 6-hourly time stamps (00:00, 06:00, 12:00, 18:00 UTC) and evaluate the 00Z and 12Z initializations and do not apply any intensity filter. We note that the definition of TC intensity is inconsistent across different meteorological agencies~\citep{schreck2014impact}. In TCBench, we use the U.S. agencies' definition: minimum sea-level pressure and 1-minute maximum sustained wind speed at an altitude of 10 meters, both of which are provided in IBTrACS. Another source of uncertainty and inconsistency across agencies and time periods is the first data point included in each TC track in the IBTrACS dataset, e.g. there is no specific intensity for when the agencies start the track of a specific TC. We provide tools for processing the IBTrACS file provided by NOAA and provide the subsection of data of interest per identified TC.%Figure \ref{fig:datasets}a,b show examples of IBTrACS data--Figure \ref{fig:datasets}a shows the tropical cyclone tracks from the 2023 season, with intensity represented by line color; while Figure \ref{fig:datasets}b provides a gridded estimate of TC count by location for the TCBench training and validation periods (2017-2022). 

ERA5~\citep{hersbach2020era5} is the fifth generation reanalysis product provided by the European Centre for Medium-Range Weather Forecasts (ECMWF). ERA5 provides dozens of meteorological variables at high spatial, vertical, and temporal resolutions, and models or assimilates a large amount of historical data. Though ERA5 is used to provide initial conditions to physical and neural weather models, we do not provide this data as it is made available by the ECMWF through its Copernicus programme.
%%%%%%%%%%%%%%%%
%\subsection{Numerical and AI Weather Prediction Model Data}

Physics-based models predict the weather by solving partial differential equations, such as the laws of fluid dynamics, thermodynamics, radiative transfer, and atmospheric chemistry. We include hindcasts from several of these models via The International Grand Global Ensemble (TIGGE) product, including ensemble hindcasts of the Global Ensemble Forecasting System (the GEFS) and the Intergrated Forecasting System (the IFS)\citep{Section4_Bougeault_tigge}. In contrast, neural weather models make forecasts using entirely data-driven methods. These methods vary by model but include graph neural networks~\citep{Section4_lam2023graphcast}, Fourier neural operators~\citep{Section4_pathak2022fourcastnet}, and vision transformers~\citep{Section4_bi2023panguweather}. We include hindcasts from several neural weather models, including NVIDIA's FourCastNetV2~\citep{Section4_bonev2023spherical}, Huawei's Pangu-Weather~\citep{Section4_bi2023panguweather}, Google DeepMind's GenCast~\citep{price2025probabilistic} and FNV3P2~\citep{alet2025skillfuljointprobabilisticweather,weatherlabteam2025tropicalcycloneai, deepmind_weatherlab}, and the ECMWF's single-member AIFS v1.0~\citep{lang2024aifs}.

All datasets were reformatted into a unified structure based on IBTrACS for consistency across sources, using the IBTrACS storm identifier for identifying TCs and following unit conventions in IBTrACS. For the physics-based models, track data (position, intensity) from international weather prediction centers was extracted from \texttt{.xml} files and standardized to a common spreadsheet format that includes timestamps and the predicted values (detailed in Table S5 in the Appendix).
%Gridded data (e.g., neural weather model outputs) was converted into storm-based tracks, using the IBTrACS storm paths as a reference. 
For neural weather model forecasts, forecast tracks and intensities were parsed from model outputs and matched to IBTrACS using TempestExtremes~\citep{tempestextremes2021} and HuracanPy~\citep{Bourdin2025}, bias corrected by dividing the winds by 0.88 following~\cite{knapp2010international}, and then standardized to the aforementioned spreadsheet format. The post-processing models we provide predict intensification via a parametric distribution, and we thus sample in order to generate a spreadsheet that follows the standardized format. Additional details for the post-processing models are provided in Appendix~D.4.
%%%%%%%%%%%%%%%%
\subsection{Data Inclusion Criteria} \label{sec:inclusion_criteria}
To ensure consistency and usefulness across all experiments, each forecast or model dataset included in TCBench had to meet the following requirements: a) They must provide at least two forecast initializations per day (00Z, 12Z); b) They must provide forecasts at least through a five day lead time, or forecasts until the IBTrACS record for that TC ends; c) the forecasts must be available as 6-hourly forecast data; d) the forecasts must provide at least the latitude, longitude, minimum pressure, and maximum wind speed variables associated with a track; e) models must provide any accompanying environmental fields (e.g., surface wind speed, surface pressure, geopotential height%\rev{
; additional examples are given in Appendix C%}
) used for tracking and intensity estimation; f) if tracking storms in environmental fields, a documented tracking algorithm and reproducible code examples; and g) if not directly forecasting track and intensity, the provided fields must have a minimum resolution %\rev{
corresponding to about
%} 
0.5° %\rev{
on a global lat-lon grid%}
, preferably 0.25°. These criteria are guidelines for both the data we provide, and data provided by community members in the future.
% \begin{addmargin}[1em]{0em}
%     $\bullet\;$ At least 2 forecast initializations per day (00z, 12z);\vspace{\listspacing}\\
%     $\bullet\;$ Minimum 5-day lead time, or forecasts until the IBTrACS record for that TC ends;\vspace{\listspacing}\\
%     $\bullet\;$ Availability of 6-hourly forecast data;\vspace{\listspacing}\\
%     $\bullet\;$ Track variables: latitude, longitude, minimum pressure, and maximum wind speed;\vspace{\listspacing}\\
%     $\bullet\;$ Accompanying environmental fields (e.g., surface wind speed, surface pressure, geopotential height) used for tracking and intensity estimation; \vspace{\listspacing}\\
%     $\bullet\;$ Documented tracking algorithm and reproducible code examples;\vspace{\listspacing}\\
%     $\bullet\;$ If not directly forecasting track and intensity: Minimum resolution of 0.5°, preferably 0.25°.\vspace{\listspacing}\\
% \end{addmargin} \vspace{-1.5em}
% \begin{itemize}
%     \item At least 2 forecast initializations per day (00z, 12z);
%     \item Minimum 5-day lead time, or forecasts until the IBTrACS record for that TC ends;
%     \item Availability of 6-hourly forecast data;
%     \item Track variables: latitude, longitude, minimum pressure, and maximum wind speed;
%     \item Accompanying environmental fields (e.g., surface wind speed, surface pressure, geopotential height) used for tracking and intensity estimation; 
%     \item Documented tracking algorithm and reproducible code examples;
%     \item If not directly forecasting track and intensity: Minimum resolution of 0.5°, preferably 0.25°.
% \end{itemize}

\subsection{Guidelines for Model Submission}

Dataset Split Recommendation:
IBTrACS storms are divided based on the year they occurred to prevent leakage, noting that we divide based on calendar year and not TC season. We suggest at least four years be used for training and at least two years be used for validation, noting that we here assign 2017-2020 to training and 2021-2022 to validation. We require that the year 2023 be left for testing, given that the models we include as baselines are not trained or tuned on this year.
 %%%%%%%%%%

Model Submission Guidelines:
TCBench submissions must include: a) a list of open data sources used for model training, including any relevant links; b) any additional data needed to run the model on the test set; c) the minimal data sources needed to train the baselines presented for the training and validation years, d) code and hyperparameters to train and run the model and generate the predictions, e) the model and/or trained weights in a standard format, f) and an environment file to recreate the computing environment (or a container with which to run the model).

All data will be released as appropriate.
The provided tracks should be included as a Comma Separated Value (csv) file following the schema outlined in Table S5, and must include the \textit{sid}, \textit{initial\_time}, and \textit{valid\_time} columns as these are used as the keys for each prediction during all metric evaluations. For probabilistic models, the provided csv should further include an \textit{ensemble\_idx} column that provides a numerical identifier for each individual ensemble member. Maximum sustained winds should be reported as 1-min maximum sustained wind in knots, minimum sea-level pressure should be reported in hPa, latitude in degrees ($ -90^{\circ} \leq \mathrm{lat} \leq 90^{\circ}$), and longitude in degrees ($0^{\circ}\leq\mathrm{lon}<360^{\circ}$).

\section{TCBench Benchmarking Toolbox}
TCBench provides a standardized set of metrics to provide objective comparisons for TC forecasting. 

Deterministic evaluation of track and intensity: We evaluate intensity forecasts using the root mean square error (RMSE) and the mean absolute error (MAE). %, and the \textbf{coefficient of determination (R$^2$)}. 
To evaluate predicted storm tracks, we compute three deterministic metrics following~\citep{heming2017tropical}: \mbox{(1) The} Direct Positional Error (DPE), which is the distance between the forecast and observed storm positions at the same verification time;  \mbox{(2) the} Cross-Track Error (CTE), which is the component of the positional error perpendicular to the observed storm motion; and \mbox{(3) the} Along-Track Error (ATE), which is the component of the positional error parallel to the observed storm motion. A complete description of the formulae and computational procedures is provided in Appendix~E.

Probabilistic evaluation of track and intensity: For probabilistic models, we consider predictions to have been made when 30\% of all ensemble members provide a value (i.e., by considering predictions only when the number of ensemble members with a prediction for the track, intensity, and/or RI divided by the total number of ensemble members $\geq$ 0.30). To quantify uncertainty in ensemble forecasts, we report the fair variant of the continuous ranked probability score (CRPS), noting that we refer to the fair CRPS whenever we mention the term in the main article. Given an ensemble of $N$ forecast values $\{x_i\}_{i=1}^N$ and an observation $y$, the CRPS provides an unbiased estimation of the discrepancy between the forecast distribution and the observed outcome~\citep{zamo2018estimation}:
\begin{equation}
\mathrm{CRPS}(\{x_i\}, y) = \frac{1}{N} \sum_{i=1}^N |x_i - y| - \frac{1}{2N\left(N-1\right)} \sum_{i=1}^N \sum_{j=1}^N |x_i - x_j|.
\label{eq:CRPS}
\end{equation}
The first term captures ensemble accuracy (mean absolute error), while the second reflects ensemble sharpness (spread). Lower CRPS indicates better-calibrated forecasts centered on the observation. Following standard practice of TC forecasting centers~\citep{narita2015probability}, track uncertainties at different temporal horizons are illustrated with probability circles drawn at the center of the ensemble mean with a radius corresponding to the n\textsuperscript{th} quantile of the distance between each ensemble member and the mean position of the remaining ensemble members---noting that standard practice utilizes the 0.70 quantile. We summarize the performance of probabilistic models by reporting the hit ratio when compared to observations (i.e.,~the percentage of times that the observed storm falls within the circle drawn with the 0.70 quantile radius), noting that a well calibrated model should have a hit ratio of $\approx 70\%$ ( $>0.70$ values correspond to overdispersive, under-confident models; conversely, $<0.70$ values indicate underdispersive, over-confident models).

Classification based evaluations: We evaluate models on the task of predicting RI using confusion matrix based metrics, including the Critical Success Index (CSI)---also known as the Threat Score---and the Peirce Skill Score (PSS). Though RI constitutes a rare event, it is not exceedingly so; we define it as approximately the 95\textsuperscript{th} percentile of intensity change over a 24-hour window. Thus, we evaluate model performance on RI forecasts using metrics more tailored for rare event forecasts. We note that while most of TCBench is configured as a regression problem, we have formulated RI as a binary classification problem. Thus, RI models are tasked with making a simple ``yes/no" prediction for the occurrence of RI (i.e., the presence of an intensification of 30 knots per 24h window, where the window is rolled over the 120h forecasts). 
For these metrics, TP refers to true positives (hits); TN refers to true negatives, or correct negatives; FP refers to false positives, or false alarms; and FN refers to false negatives, or misses. 
% Classification Metrics used for RI in our case
The CSI measures the ratio of correctly predicted positive observations to the sum of all predicted positives, actual positives, and minus true positives. 
% CSI can be used both for probabilistic track evaluation and RI. 
\[
    \mathrm{CSI} = \frac{\mathrm{TP}}{\mathrm{TP} + \mathrm{FN} + \mathrm{FP}}
\]
The PSS measures the relation between the True Positive Rate and False Positive Rate.
\[
    \mathrm{PSS} = \mathrm{TPR} - \mathrm{FPR} = \frac{\mathrm{TP}}{\mathrm{TP} + \mathrm{FN}} - \frac{\mathrm{FP}}{\mathrm{FP} + \mathrm{TN}}
\]
We further provide the Brier Skill Score (BSS) as a metric for evaluating the RI forecasts. The BSS compares the Brier score of a model to a reference, where the Brier score is the mean of the difference between the predicted and observed probability of RI: $\sfrac{1}{N}\sum_{i=1}^{N}{\left( p_i - o_i \right)^2}$. For deterministic models, we define the predicted probability as 1 or 0 given TCBench expects a binary value as an RI prediction; for ensembles, we use the fraction of all ensemble members predicting RI. For the BSS, we use the mean tendency by lead and basin as the reference model, which is further described in the Baselines section.

\section{Baselines}

Following standard evaluation procedures~\citep{Section2_Knaff_persistence2003}, we first compare TC forecasts to the Mean Tendency by Lead \& Basin (CLIM) and persistence baselines. CLIM uses the mean IBTrACS changes in intensity and position relative to the initial conditions, calculated from 1980 to 2022, conditioned on basin and lead time. For persistence, we predict subsequent time steps assuming a constant intensity and a constant vector of motion across the earth surface given the state of the storm at the initial time of forecast (i.e., we predict intensity to remain constant and storm motion to evolve at the same rate for all time steps).

Baselines for Track and Intensity Predictions:
We provide scores for tracks obtained directly from DeepMind's Weather Lab website for GenCast and WeatherNext 2 Cyclones~\citep[FNV3P2,][]{Alet2026}, and for the ECMWF IFS ensemble and the NCEP GEFS models using tracks from the TIGGE dataset.
%As baselines for track predictions, we process global outputs from 
For Pangu-Weather, FourCastNetv2, and the single-member version of AIFS, we process global outputs to obtain tracks using TempestExtremes~\citep{tempestextremes2021}. The parameters we used are listed in \Cref{tab:tempextremes_combined}, and are the same as in~\citep{tempestextremes2021}, with the exception of the minimum track duration, which is set to 12h. While this would normally lead to overly sensitive tracking (i.e., resulting in less filtering of short-lived systems that are not true cyclones), we frame the problem of TC prediction with the assumption that there is a prior system of interest known to exist at an initial time and we are interested in the evolution of said tropical system. We thus eliminate all spurious tracks that could not be matched to the systems of interest. 

To ensure a fair comparison between model predictions and best tracks, only the tracks that have direct temporal and spatial correspondence to the best track will be retained for subsequent analyses. Cases that are extraneous to an observed storm at the time of forecast (i.e., false alarms) are excluded from the evaluations, and prediction misses are further excluded from the calculations. Additionally, we also remove early and late dissipation tracks for metric calculations---referring to prediction tracks that are shorter and longer than the official best tracks, respectively. 

% Avoiding former/latter formulation not to confuse readers

%For GenCast and Weatherlab FNV3, we also provide scores for the tracks made available by DeepMind through the Weather Lab website. Finally, we also provide scores for the ECMWF IFS ensemble and the NCEP GEFS dynamical weather models as represented by the tracks provided in the TIGGE dataset. 

% \begin{table}
%     \centering
%     \begin{tabular}{cccc}
%          Variable \hspace{.5cm}&  Delta \hspace{.5cm}& Dist. ($^\circ$) \hspace{.5cm} & Min/Max Dist. ($^{\circ}$) \\
%          \hline
%          MSLP& 200 Pa & 6.5 & 0\\
%          $\Delta z_{300,500}$& -58.8 $m^2s^{-2}$ & 5.5 & 1.0\\
%     \end{tabular}
%     \caption{TempestExtremes Closed Contour Command Parameters. Distances are expressed in great circle degrees.}
%     \label{tab:tempextremes_contour}
% \end{table}

% \begin{table}
%     \centering
%     \begin{tabular}{lc}
%          Parameter \hspace{.5cm}&  Value \\
%          \hline
%          Minimum Track Duration & 12h \\
%          $V_{\mathrm{max}}$ Threshold& 10 m/s\\
%          Latitude ($\phi$) & $\left| \phi \right| \leq 50$
%     \end{tabular}
%     \caption{TempestExtremes Stitch Nodes Parameters. Distances are expressed in great circle degrees.}
%     \label{tab:tempextremes_stitch}
% \end{table}

\begin{table}
    \centering
    \small
    \setlength{\tabcolsep}{4pt}
    \renewcommand{\arraystretch}{1.15}
    \begin{tabular}{@{}lccc@{\hspace{1em}}lc@{}}
    \toprule
    \multicolumn{4}{c}{\textbf{Closed Contour Command Parameters}} & \multicolumn{2}{c}{\textbf{DetectNodes \& StitchNodes}} \\
    \cmidrule(lr){1-4}\cmidrule(l){5-6}
    \textbf{Variable} & \textbf{Delta} & \textbf{Dist. ($^\circ$)} & \textbf{Min/Max Dist. ($^\circ$)} & \textbf{Parameter} & \textbf{Value} \\
    \midrule
    MSLP & 200 Pa & 5.5 & 0 & Merge Distance & 6.0$^\circ$ \\
    $\Delta z_{300,500}$ & -58.8 $\mathrm{m}^2\mathrm{s}^{-2}$ & 6.5 & 1.0 & Radius $V_{max}$ ($R_1$) & 2.0$^\circ$ \\
     &  &  &  & StitchNodes Range & 8.0$^\circ$ \\
     &  &  &  & Min. Track Duration & 12h \\
     &  &  &  & $V_{max}$ Thresh. & $\geq$10 m/s ($\geq$2 nodes) \\
     &  &  &  & Latitude ($\phi$) & $|\phi| \leq 50^\circ$ ($\geq$1 node) \\
    \bottomrule
    \end{tabular}
    \caption{TempestExtremes Parameters. Distances are expressed in great circle degrees. $R_1$ is the wind-extraction radius. Unlisted parameters use the default parameter values present in TempestExtremes.}
    \label{tab:tempextremes_combined}
\end{table}

We provide additional baselines that post-process data-driven weather model forecasts to predict tropical cyclone intensity, using a pipeline that follows~\citep{gomez2025global}. These baselines provide only an intensity forecast, and a more thorough description of these baselines is given in Appendix~D.4.
Finally, we use all baselines for intensity prediction to evaluate for RI by thresholding  the wind speed change---i.e., if the 24 hour change in wind speed is at least 30 kts, RI has occurred~\citep{kaplan2003,Kaplanetal2010}. Shorter  lead times partially include observed IBTrACS winds preceding the forecast initialization, which can contribute to non-zero RI skill for persistence and climatology and inflate short-lead scores. 

\section{Results}

\begin{figure}[h]
    \centering
    \includegraphics[width=\linewidth]{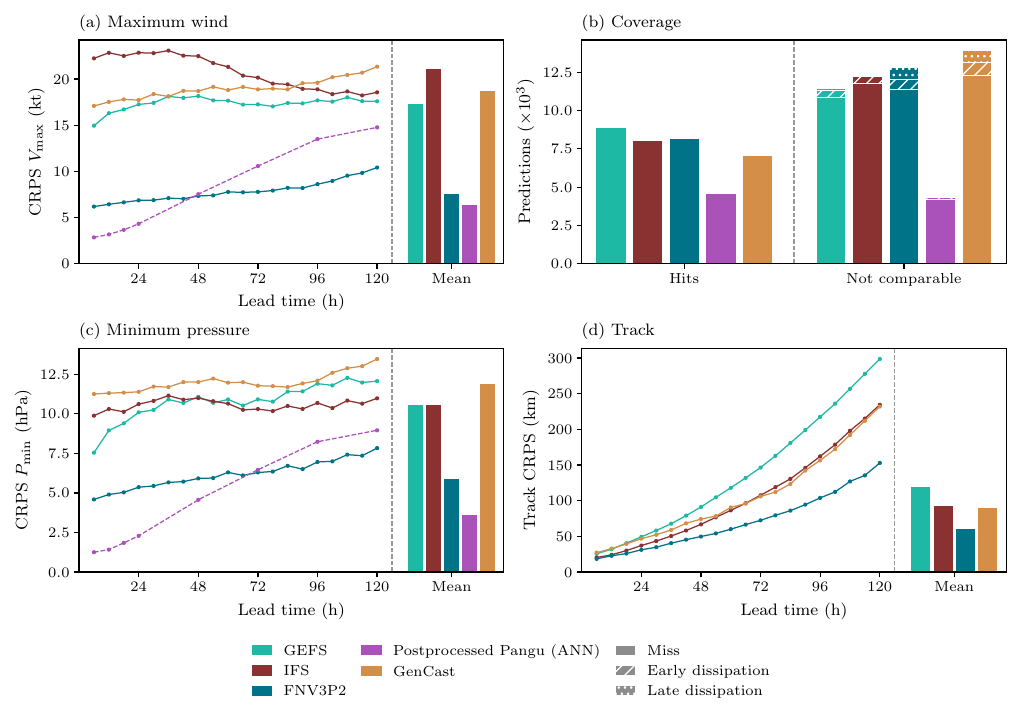}
    \caption{Summary of probabilistic model performance on the 2023 set, made with TCBench. Panel (b) shows a count of model predictions for which an observation exists with which to compare against (hit) on the left, while the right shows a count of timestamps that cannot be compared numerically (misses \& early dissipation for timestamps for which the models produce no prediction, late dissipation for timestamps for which the models provide a prediction but no corresponding observation exists). Each metric panel shows a line plot that breaks down the metric per lead time, and a mean metric bar plot is shown on the right of each panel.}
    \label{fig:probabilistic_summary}
\end{figure}

\begin{figure}[h]
    \centering
    \includegraphics[width=\linewidth]{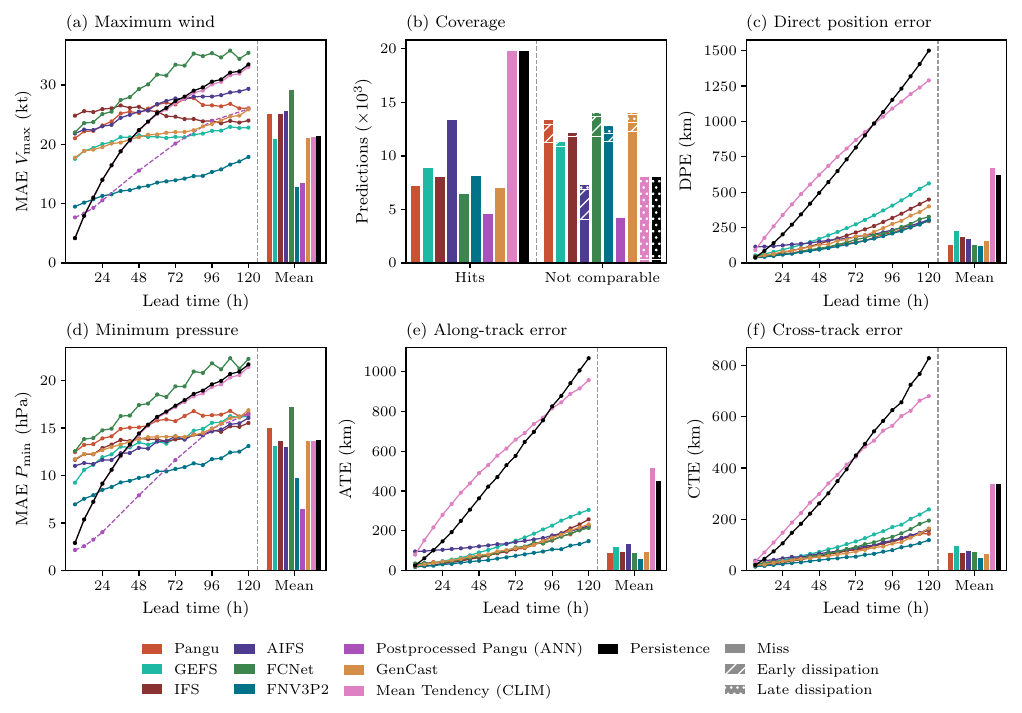}
    \caption{Summary of deterministic model performance on the 2023 set, made with TCBench. Panel (b) shows a count of model predictions for which an observation exists with which to compare against (hit) on the left, while the right shows a count of timestamps that cannot be compared numerically (misses \& early dissipation for timestamps for which the models produce no prediction, late dissipation for timestamps for which the models provide a prediction but no corresponding observation exists). Each metric panel shows a line plot that breaks down the metric per lead time, and a mean metric bar plot is shown on the right of each panel.}
    \label{fig:deterministic_summary}
\end{figure}

In this section we present examples of using TCBench to evaluate the deterministic and probabilistic skills of neural weather models (AI Models) and physics-based  models on TC track and intensity predictions, covering lead times of up to five days. The results are summarized in \Cref{fig:deterministic_summary,fig:probabilistic_summary}, noting that in this figure we evaluate models only for cases with a direct temporal match between model predicted tracks and the best track. Additionally, we select the MAE as the summary metric for the deterministic panel given that the CRPS generalizes the MAE to probabilistic forecasts. For all probabilistic models, an ensemble fraction of 0.30 is used to determine whether the ensemble predicts the occurrence of a TC, and deterministic evaluations use the averaged errors of ensemble members predicting a value whenever a deterministic evaluation metric is calculated. % and Figure \ref{fig:fair-comparison}.
%We evaluate the performance of NWP and AI models used in TCBench for both track and intensity predictions using the deterministic (DPR, MAE) and probabilistic (CRPS) metrics over lead times up to five days. The results are summarized in Figure~\ref{fig:schematic} and Figure \ref{fig:fair-comparison}.

\textbf{Most models show significant room for improvement with respect to coverage.} In panel (b) shown in \Cref{fig:deterministic_summary,fig:probabilistic_summary}, the persistence and mean tendency baselines show the upper limit for covering all cases. When comparing the models evaluated in TCBench, we observe that only AIFS surpasses a 50\% coverage rate. This is likely indicative of models generally struggling with lower intensity systems, a known issue associated with the relatively coarse resolution of the models and known issue with the ERA target training data for the AI models~\citep{bourdin2022}.

\textbf{Some AI models perform well for tracks:} 
As shown in panels (c), (e), and (f) of \Cref{fig:deterministic_summary} and panel (d) of \Cref{fig:probabilistic_summary}, FNV3P2 and GenCast tend to perform well on their predicted tracks, the former exhibiting the lowest errors across most lead times and the latter performing on par with IFS, a trusted traditional NWP model. Furthermore, all of the tested models significantly outperform simple baselines across most lead times---confirming that the models provide value in modeling tropical cyclone trajectories. 

\textbf{Task-specific adjustments to AI models (post-processing or otherwise) yield skillful intensity predictions:} The coarse spatial resolution of the physics-based global weather prediction models and the underlying training data for the AI models mean that neither type of model is likely to perform well for intensity predictions. In \Cref{fig:deterministic_summary} we see models not only produce forecasts of $V_{\mathrm{max}}$ and $p_{\mathrm{min}}$ that are worse than the persistence and mean tendency baselines for shorter lead times but also outperform them for longer lead times (e.g., lead times $\geq$ 72h), except FCNet. We see that at very short lead times performance is heavily influenced by assumptions made on what information is available at the initial time of forecast (e.g., the position and intensity ascribed by TCVitals~\cite{tcvitals}).

%As shown in \Cref{fig:fair-comparison}, track prediction skills deteriorate with increasing lead time. All evaluated models exhibit track errors that are smaller than the persistence error at all lead times, indicating that all models produce useful track predictions. This is especially apparent when looking at the error for the predicted tracks without a fallback to persistence, shown by Figure S2 in Appendix~F. % TODO
%While the track error for most AI models and physics-based models are very similar across lead times, we see a clear indication of the AIFS model to outperform other models in deterministic metrics beyond 24 hours. 
% We largely attribute this to AIFS producing TC track predictions for a greater number of time steps (see Figure S3 in Appendix~F~), which greatly reduces the number of samples for which the evaluation falls back to the persistence baseline. Regarding the quality of model track ensembles, the physics-based GEFS model still outperforms the AI model alternatives we evaluated with regards to track displacement CRPS. These results highlight the potential of some existing data-driven forecasting models to provide reliable deterministic TC track forecasts, but suggest that they are yet to be as good as physics-based ensemble models in probabilistic track predictions.

It is important to note that the physics-based GEFS continues to be among the best performing models, both in terms of error magnitude and of coverage. Additionally, we find that a well-designed post-processing algorithm of raw AI predictions can improve upon an originally inaccurate AI model (e.g., Pangu-Weather) to the extent of producing predictions on par with or better than significantly more sophisticated models. The results for the post-processed models are, however, positively biased as they were initiated with ERA5 data as the models have not been fine-tuned on operational conditions (noting that the versions of FourCastNetV2, and Pangu-Weather we use were also all initialized on ERA5 data for the same reason, unlike AIFS which was initialized with the operational analysis stream made available on MARS by the ECMWF). The results further show a clear advantage by models that are trained specifically for predicting extreme events (e.g., FNV3P2), as they can significantly outperform models that are trained only to predict global weather.

%However, the difference in deterministic and probabilistic skills between the best raw AI model and GEFS is slight. Furthermore, we do not discount the possibility that a high spatial resolution, non-global physics-based model can be dramatically better than both GEFS and the AI models at predicting TC intensity. We find that a well-designed post-processing algorithm of raw AI predictions can improve upon an originally inaccurate AI model (e.g., PanguWeather) to the extent of producing predictions on par with or better than the GEFS model. Additionally, it is clear that models that are trained specifically for predicting extreme events (e.g., FNV3) can significantly outperform models that are trained only to predict global weather. 

\begin{figure}[h]
  \centering
  \includegraphics[width=1\textwidth]{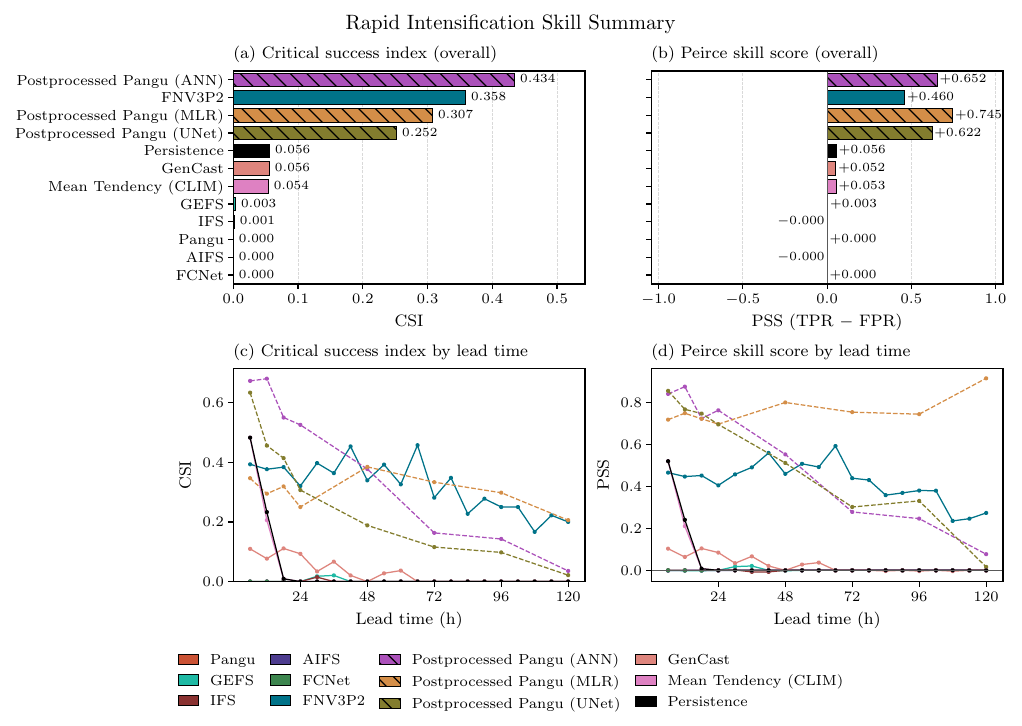}
  \caption{Rapid Intensification Summary. Hatching and dashed lines denote models that do not output predictions for all lead times evaluated in TCBench.}
  \label{fig:ri_summary}
\end{figure}

%\paragraph{
\textbf{Postprocessed AI models and FNV3P2 capture RI:} The challenges of predicting TC RI events with existing global AI and physics-based models can clearly be observed in ~\Cref{fig:ri_summary}, where most models show little to no ability to forecast such events across different lead times. Of the models we evaluated, FNV3P2 and the post-processed Pangu-Weather model show the most meaningful RI skill, although GEFS and GenCast exhibit small positive BSS values at some lead times. The lead-time dependence of RI BSS, together with the intensity RMSE comparisons against persistence, is summarized in ~\Cref{fig:deterministic_scorecard}. We note that neural weather models have generally been trained on ERA5 reanalysis, which is known to have a negative bias with respect to TC intensity. 

This issue, combined with the fact that many models have been trained to reduced mean error (e.g., MAE, RMSE) and the relatively coarse resolution of the predicted fields, makes predictions of RI by the models a significantly more challenging task. Additionally, we expect that the performance of the post-processing models is positively biased due to the construction of the input processing pipeline, which was developed for research purposes and does not generate the inputs for dates where the observed storm dissipates. We expect that the false positive rate is higher than what is observed in the provided tracks, and given that the number of false positives would make the CSI and PSS scores lower, some amount of skill degradation is expected. 
Overall, our evaluation reveals that: 
\begin{itemize}%[leftmargin=1em]
    \item There is room for improvement on prediction coverage across traditional NWP and AIWP models, and the ECMWF's AIFS seems to be the model with the widest breadth of predictions.
    \item GEFS is amongst the best performing models, but FNV3P2 largely outperforms other models on the intensity predictions.
    \item Intensity forecasts from AI models can be improved with post-processing or task-specific training.\looseness=-1
\end{itemize}

%     \item Some neural weather models outperform GEFS in deterministic track prediction, but not probabilistic ones.
%     \item GEFS provides slightly better deterministic and probabilistic intensity forecasts than AI models.
%     \item Intensity forecasts from AI models can be improved with a post-processing pipeline.
%     %\item TIGGE GEFS probabilistic forecasts exhibit appropriate error growth but are under-dispersive at extended lead times.
% \end{itemize}
These findings emphasize the complementary strengths of neural and physics-based systems, and point toward hybrid approaches as a promising future direction in TC forecasting. These findings also agree with~\citep{demaria2024evaluationtropicalcyclonetrack,sahu2025TCevaluation}, who found that neural weather models could make skillful TC track predictions, but lacked skill in intensity forecasting. The results from the post-processed neural models are encouraging and point the way towards making data-driven TC intensity forecasts more reliable.

\begin{figure}[h]
  \centering
  \includegraphics[width=1\textwidth]{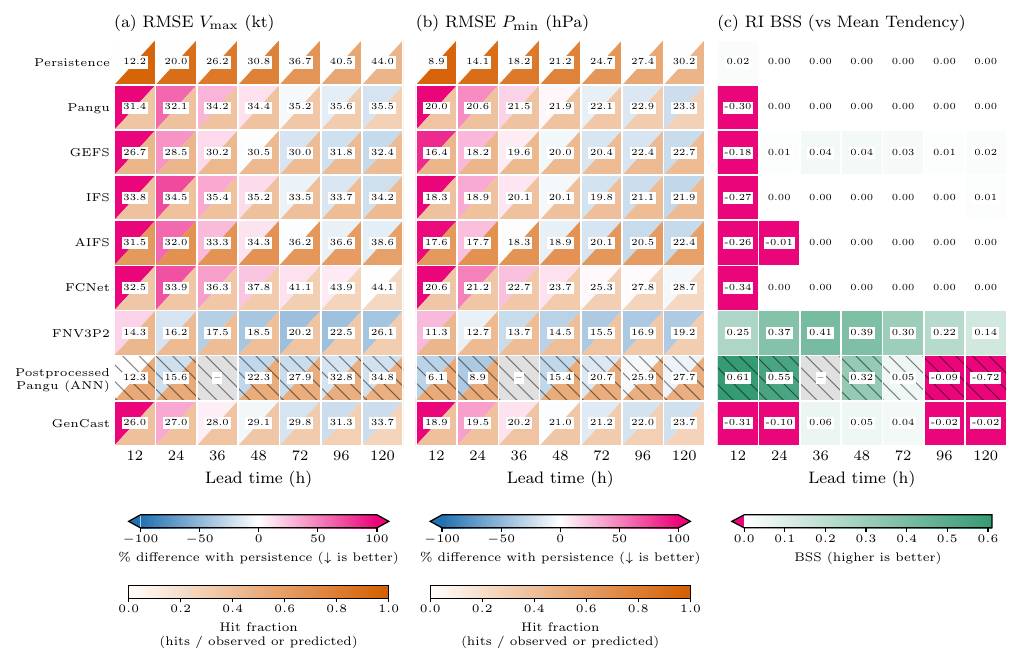}
  \caption{Deterministic and RI Scorecard - the cards showcase the Root Mean Squared Error (RMSE) and Brier Skill Score (BSS)---compared against the mean tendency by lead time and basin---for a subset of predicted lead times. Hatched rows indicate models that do not predict all 6-hourly lead times up to 5 days.}
  \label{fig:deterministic_scorecard}
\end{figure}

% \begin{figure}[h]
%   \centering
%    \includegraphics[width=1\textwidth]{Evaluation_Figures/ScoreCard_Deterministic.pdf}
%   \caption{TCBench deterministic scorecard (2023, baseline = persistence).
% Cells show mean error; colors show \% difference vs baseline at the same lead (12-120h; columns) for each model (rows).}
%   \label{fig:deterministic_scorecard}
% \end{figure}

% \section{Other Potential Applications, Limitations, and Conclusion}
\section{Discussion and Conclusion} % More complete alternative above

%\paragraph{Other Potential Applications:} 
In addition to the simple deterministic and probabilistic TC track and intensity predictions presented above, the TCBench dataset can be applied to other scientific applications related to different aspects of tropical cyclones. These include assessments of wind-related risks posed by tropical cyclones, tropical cyclone wind and precipitation nowcasting, wind field reconstruction, and data aggregation for physics discovery. These potential applications are further elaborated upon in  Appendix~G.

%\begin{enumerate}
%    \item Imputation/Reconstructing fields from partial observations
%    \item Nowcasting
%    \item Downscaling
%    \item Predicting TC activity across timescales (short-term to seasonal)
%    \item Estimates of TC activity in different climates
%    \item Developing predictive theory in tropical meteorology
%\end{enumerate}

%\paragraph{Limitations:} 
There are clear limitations regarding this work, including the variance and evolution of uncertainty in the data, a limited selection of sources for tractability, and TC track inconsistencies across agencies. We acknowledge these and expand on them further in Appendix~H.
By emphasizing flexibility in data experimentation, TCBench provides a data foundation that can be extended with advancements in TC observations and modeling. TCBench's goal is to help the research community improve TC forecasts, our understanding of the physical processes governing TC behavior, ultimately mitigating TC impacts on local communities and society at large. We additionally provide a common point of comparison between TC forecasting models that is centered around the prediction of TC behavior once systems have formed. We show that for this task there is ample space for further research and improvement beyond our current data-driven forecasting capabilities, and provide tools for helping researchers in this field.

\FloatBarrier
\section{Data availability statement}
% \data{
TCBench uses the International Best Track Archive for Climate Stewardship (IBTrACS; NOAA) as ground truth for tropical cyclone track and intensity. Reanalysis data from ERA5 (ECMWF Copernicus Climate Data Store) are referenced for environmental fields but not redistributed. Physics-based ensemble forecasts are obtained from the TIGGE archive, including GEFS (NCEP) and IFS (ECMWF). Neural weather model forecasts include FourCastNetV2 (NVIDIA), Pangu-Weather (Huawei), GenCast, FNV3P2 (Google DeepMind), and AIFS v1.0 (ECMWF). Storm tracking and post-processing are performed using TempestExtremes and HuracanPy. All datasets are reformatted into a standardized IBTrACS-aligned structure and made available via the TCBench GitHub \url{(https://doi.org/10.5281/zenodo.23801372)} and HuggingFace \url{(https://doi.org/10.57967/hf/8882)} repositories.
% }

\section{Acknowledgments}
 % \ack{
 We thank Quoc-Phi Duong, Adrien Colomb, Linus Magnusson, Matthew Chantry, Jesper Dramsch, and Dan Assouline for guidance, and Margot Sirdey and Flavio Calvo for support with the local HPC system. We acknowledge the use of ERA5, AIFS, IFS analysis initialization data, and AI models from ECMWF; IBTrACS from NOAA; FourCastNetV2 from NVIDIA; Pangu from Huawei; and GenCast from Google DeepMind. We also thank the developers of \texttt{numpy}, \texttt{xarray}, \texttt{scikit-learn}, \texttt{pandas}, \texttt{torch}, \texttt{zarray}, \texttt{cuda}, \texttt{huggingface}, \texttt{TempestExtremes},  \texttt{huracanpy}, and \texttt{troPYcal}. Collaborations during the TROPICANA program (Institut Pascal at Universite Paris-Saclay, ``Investissements d'avenir'' ANR-11-IDEX-0003-01) contributed to this work. CL and SJC acknowledge the support from NASA MAP program (80NSSC21K1525).
 % }
% \funding{
SB is supported by the NERC large grant n° NE/W009587/1 (HURACAN). MM is supported by the National Science Foundation AI Institute Grant ICER-2019758. TB acknowledges support from the Swiss National Science Foundation (SNSF) under Grant No. 10001754 (``RobustSR'' project).
% }

\section{Author contributions}
% \roles{
\textbf{Milton Gomez:} Conceptualization, Methodology, Writing, Data Curation, Software, Visualization. \textbf{Saranya Ganesh S.:} Conceptualization, Writing, Visualization. \textbf{Marie McGraw:} Conceptualization, Writing, Software. \textbf{Frederick Iat-Hin Tam:} Conceptualization, Writing, Visualization. \textbf{Ilia Azizi:} Writing. \textbf{Samuel Darmon:} Software, Visualization. \textbf{Monika Feldmann:} Conceptualization, Writing. \textbf{Stella Bourdin:} Software. \textbf{Louis Poulain-$\,$-Auzéau:} Data Curation, Software. \textbf{Suzana J. Camargo:} Conceptualization. \textbf{Jonathan Lin:} Validation. \textbf{Dan Chavas:} Conceptualization. \textbf{Chia-Ying Lee:} Validation. \textbf{Ritwik Gupta:} Writing. \textbf{Andrea Jenney:} Writing. \textbf{Tom Beucler:} Conceptualization, Writing, Supervision, Project administration, Funding acquisition.
% }

\section{LLM Usage Disclaimer}
Conversational AI tools (e.g., ChatGPT and Copilot) were used for the following tasks during the manuscript's preparation: text editing (specifically, in making certain sections more succinct), LaTeX formatting, as an aid in code development when writing the benchmark framework, and to check that the text faithfully described the code. 

%
% Each of the commands below will create an unnumbered section with the appropriate heading.
% Remove any sections that are not relevant for your article.
% All sections except suppdata will be removed if the [anonymous] option is used.
% See iopjournal-guidelines.pdf for more information.
%

\bibliographystyle{unsrt}
\bibliography{references, si_references}
% 

%TODO : Figure out how to present the funding, contributions, etc,

% \ack{Sample text inserted for demonstration.}

% \funding{Sample text inserted for demonstration.}
% % This section is a list of funder names and grant numbers

% \roles{Sample text inserted for demonstration.}
% % List author names and the contributions made to the article, using terms from the NISO Contributor Roles Taxonomy (CRediT) https://credit.niso.org

% \data{Sample text inserted for demonstration.}
% % For more information on IOP Publishing's research data policy see: https://publishingsupport.iopscience.iop.org/questions/research-data/

% \suppdata{Sample text inserted for demonstration.}
% TODO: figure out supplemental information inclusion format
% \section*{Appendices}
% \appendix
\FloatBarrier
\clearpage
{
\setcounter{figure}{0}
\setcounter{table}{0}
\setcounter{equation}{0}

\renewcommand{\thesection}{\Alph{section}}
\renewcommand{\thesubsection}{\thesection.\arabic{subsection}}
\renewcommand{\thesubsubsection}{\thesubsection.\arabic{subsubsection}}
\renewcommand{\thetable}{S\arabic{table}}
\renewcommand{\thefigure}{S\arabic{figure}}
\renewcommand{\theequation}{S\arabic{equation}}

\setcounter{section}{0}
\setcounter{table}{0}

{\exhyphenpenalty=10000\hyphenpenalty=10000
\fontsize{18}{21}\selectfont\noindent\raggedright
\textsf{Supplementary Material for TCBench: A Benchmark for Tropical Cyclone Track and Intensity Forecasting at the Global Scale}\par}

\vspace{5mm}

%%%%%%%%%%%%%%%%%%%%%%%%%%%%%%%%%%%%%%%%%%%%%%%%%%%%%%%%%%%%
\section{Accountability and Reproducibility}

TCBench is released under the open-source MIT License. Continued development, including updates discussed in the limitations section, will be managed on the official TCBench GitHub page. Maintenance includes community support, issue tracking, and ongoing curation of benchmark content. All assets are hosted on GitHub, HuggingFace, and Zenodo to ensure stability and accessibility.

\vspace{0.5em}
\noindent\textbf{Resources:}
%%%%Please update all the links to ours for hugging face
% \data{
\begin{itemize}
    \item \textbf{Dataset:} Available at \url{https://doi.org/10.57967/hf/8882}\\
    \item \textbf{Model Checkpoints:} \url{https://doi.org/10.5281/zenodo.20323207}\\
    \item \textbf{Code:} \url{https://doi.org/10.5281/zenodo.23801372}\\
    \item \textbf{Container:} \url{https://doi.org/10.5281/zenodo.23759605}\\
\end{itemize}
% }
% \begin{itemize}
%     \item \textbf{Dataset:} Available at \textit{redacted for anonymity} %\url{https://huggingface.co/datasets/TCBench/TCBench}
%     \item \textbf{Model Checkpoints:} \textit{redacted for anonymity} %\url{https://huggingface.co/datasets/TCBench/TCBench/tree/main/}
%     \item \textbf{Code:} \textit{redacted for anonymity}%\url{https://github.com/msgomez06/TCBench_0.1}
%     \item \textbf{Documentation:} \textit{redacted for anonymity}%\url{https://tcbench.github.io/README.html}
% \end{itemize}

% An anonymized subset of the code and data has been prepared and uploaded to ProtonDrive:

% \url{https://drive.proton.me/urls/SKT1FGC3KG#lIOFkFcL4xOU}

%\subsection{Repository Structure}
\subsection{HuggingFace Repository Structure}
We organize the TCBench HuggingFace repository as follows:
\begin{itemize}
    \item \texttt{2023\_IBTrACS.csv}: The subset of IBTrACS with the storms and timestamps that we use to evaluate models.\\
    \item \texttt{evaluated\_tracks/}: A collection of CSVs with the evaluated baseline tracks. The plotting\_csvs subfolder further includes the summarized metrics used for generating the plots.\\
    \item \texttt{matched\_tracks/}: A collection of CSVs with the tracks we evaluate as baselines, including AIFS, \mbox{FourCastNetv2}, Weatherlab FNv3, GenCast, Pangu Weather, TIGGE GEFS, and TIGGE IFS.\\
    \item \texttt{neural\_weather\_models/}: Raw, global neural weather model outputs for up to 5 days lead time, initialized at each timestamp associated with a storm in \textit{2023\_IBTrACS.csv} (i.e., the ground truth).\\
    \item \texttt{postprocessing\_model\_data/}: Postprocessing input data required to run the models on the test set for Pangu Weather.\\
    \item \texttt{unmatched\_tracks/}: A collection of CSVs denoting tracks found in model outputs using TempestExtremes, comprising subdirectories for AIFS, \mbox{FourCastNetv2}, and PanguWeather.
\end{itemize}

\subsection{Github Repository Structure}
We organized the TCBench Github repository as follows, with all code in the \texttt{dev/} folder:
\begin{itemize}
    \item \texttt{TempestExtremes\_example.sh} and \texttt{track\_matcher.py}: The TempestExtremes routine used to generate the tracks associated with the neural weather models, and the python routine that relies on HuracanPy to match the tracks found with TempestExtremes to the TCs in IBTrACS and convert them to the TCBench format.\\
    \item \texttt{utils/}: Collection of utilities used to process IBTrACS data and to process neural weather model data for post-processing.\\
    \item \texttt{evaluate\_tracks.py}, \texttt{metrics.py}, and \texttt{plot\_errors.py}: python scripts used to generate the persistence and climatology baseline predictions and evaluate a set of tracks (\texttt{evaluate\_tracks.py}, configured by \texttt{evaluate\_tracks\_config.yaml}), compute the metrics (\texttt{metrics.py}), and produce the summary tables and figures (\texttt{plot\_errors.py}, configured by \texttt{plot\_errors\_config.yaml}). The post-processing baselines are trained with the \texttt{postprocessing\_*} scripts, and \texttt{Getting\_started.ipynb} walks through the evaluation workflow.
\end{itemize}

% \begin{itemize}
%     \item \texttt{track\_processing/}: The TempestExtremes routines needed to generate the tracks associated with the neural weather models, and the python routines that rely on HuracanPY to match the tracks found with TempestExtremes to the TCs in IBTrACS.
%     \item \texttt{utils/}: Collection of utilities used to process IBTrACS data, to process neural weather model data for post-processing, and to calculate metrics on the tracks.
%     \item \texttt{scriptnames.py}: various python scripts used to generate the track baseline predictions (e.g.,~\textit{\mbox{baselines.py}, climatology\_maker.py, compute\_persistence.py}), to evaluate a set of tracks (e.g.,~\textit{\mbox{evaluate\_tracks.py}}) 
% \end{itemize}

% Summary of licensing, data/code availability, and long-term maintenance plans.
%%%%%%%%%%%%%%%%%%%%%%%%%%%%%%%%%%%%%%%%
\section{Getting Started}
%Add the Tutorial here; Installation; detailed Readme;
% \data{
We provide a getting started jupyter notebook that provides instructions for preparing the necessary data and benchmarking your own model on our website. Please visit the following URL to see the notebook:\\\url{https://github.com/tcbench/TCBench/blob/main/dev/Getting_started.ipynb}\\%\url{https://tcbench.github.io/README.html}
\noindent We also ask that if you encounter any issues to please feel free to contact us or raise an issue on GitHub.
% }

% A getting started notebook for the anonymized review was prepared and is available at:\\
% \url{https://drive.proton.me/urls/SKT1FGC3KG#lIOFkFcL4xOU} \\
% under the filename \textit{Getting\_Started.ipynb}

% \subsection{Data Preparation} % B.1

% First, navigate to the repository directory and install the required dependencies:

% Discuss relevant S2S or tropical cyclone benchmarks and how TCBench differs.

%%%%%%%%%%%%%%%%%%%%%%%%%%%%%%%%%%%%%%%%
\section{TCBench Data Organization and Workflow}

\subsection{Data Sources}
TCBench integrates multiple observational, reanalysis, forecast model, and AI-based data sources relevant to tropical cyclone (TC) analysis and prediction. Table~\ref{tab:data_sources} summarizes datasets currently included, and points to additional datasets of relevance. Table ~\ref{tab:list_cases} shows the IBTrACS storm identification numbers (SIDs) and basin for each tropical cyclone in the 2023 evaluation set. 

\begin{table}[ht]
    \centering
    \small
    % \begin{tabular}{p{2.5cm}p{4.25cm}p{5cm}p{1.5cm}}
    %     \toprule
    %     \textbf{Data Source} & \textbf{Description} & \textbf{Website} & \textbf{Provided} \\
    %     \midrule
    %     \multicolumn{3}{l}{\textbf{Reanalysis}} \\
    %     \midrule
    %     ERA5 & European Re-Analysis 5 & \url{https://cds.climate.copernicus.eu/datasets/reanalysis-era5-pressure-levels?tab=overview} & No \\ %\arrayrulecolor{blue}
    %     \midrule
    %     \multicolumn{3}{l}{\textbf{Operational Analysis}} \\
    %     \midrule
    %     HRES Analysis & ECMWF's archive of operational analysis, distributed through MARS & \url{https://confluence.ecmwf.int/display/UDOC/MARS+content#MARScontent-Atmosphericmodels#Analysis:~:text=models-,Analysis} & No\\
    %     \arrayrulecolor{black}
    %     \midrule
    %     \multicolumn{3}{l}{\textbf{Ground Truth}} \\
    %     \midrule
    %     IBTrACS & Best Track Archive & \url{https://www.ncdc.noaa.gov/ibtracs/} & Yes \\
    %     Extended Best-Tracks & TC tracks with size & \url{https://rammb2.cira.colostate.edu/research/tropical-cyclones/tc_extended_best_track_dataset/} & No \\
    %     TC PRIMED & ML Ready Benchmark dataset & \url{https://rammb-data.cira.colostate.edu/tcprimed/} & No \\
    %     \midrule
    %     \multicolumn{3}{l}{\textbf{Statistical Models}} \\
    %     \midrule
    %     SHIPS & Statistical Hurricane Intensity Prediction Scheme (developmental dataset) & \url{https://rammb2.cira.colostate.edu/research/tropical-cyclones/ships/development_data/} & No \\
    %     \midrule
    %     \multicolumn{3}{l}{\textbf{Models}} \\
    %     \midrule
    %     % NCEP GFS & Global Forecast System & \url{https://www.nco.ncep.noaa.gov/pmb/products/gfs/} & No \\
    %     % TIGGE GFS & NCEP GEFS ensemble model tracks dataset & \url{https://rda.ucar.edu/datasets/d330003/dataaccess/#} & Yes\textsuperscript{\textdagger} \\
    %     TIGGE IFS & ECMWF IFS ensemble model tracks dataset & \url{https://rda.ucar.edu/datasets/d330003/dataaccess/#} & Yes\textsuperscript{\textdagger} \\
    %     PanguWeather & AI global model outputs & \url{https://github.com/198808xc/Pangu-Weather} & Yes \\
    %     FourCastNet & AI model outputs & \url{https://github.com/NVlabs/FourCastNet} & No \\
    %     FourCastNetv2 & AI model outputs & \url{https://github.com/NVIDIA/torch-harmonics} & Yes \\
    %     AIFS & AI global forecasting & \url{https://arxiv.org/abs/2406.01465} & Yes \\
    %     \bottomrule
    % \end{tabular}
\begin{tabular}{p{2.5cm}p{4.25cm}p{5cm}p{1.5cm}}
    \toprule
    \textbf{Data Source} & \textbf{Description} & \textbf{Website} & \textbf{Provided} \\
    \midrule
    \multicolumn{4}{l}{\textbf{Ground truth}} \\
    \midrule
    IBTrACS & Best Track Archive & \url{https://www.ncdc.noaa.gov/ibtracs/} & Yes \\
    Extended Best-Tracks & TC tracks with size & \url{https://rammb2.cira.colostate.edu/research/tropical-cyclones/tc_extended_best_track_dataset/} & No \\
    \midrule
    \multicolumn{4}{l}{\textbf{Reanalysis and operational analysis}} \\
    \midrule
    ERA5 & ECMWF Reanalysis v5 & \url{https://cds.climate.copernicus.eu/datasets/reanalysis-era5-pressure-levels?tab=overview} & No \\
    HRES Analysis & ECMWF operational analysis (MARS) & \url{https://confluence.ecmwf.int/display/UDOC/MARS+content#MARScontent-Atmosphericmodels#Analysis:~:text=models-,Analysis} & No \\
    \midrule
    \multicolumn{4}{l}{\textbf{Physics-based global models (TIGGE)}} \\
    \midrule
    GEFS & NCEP ensemble tracks & \url{https://rda.ucar.edu/datasets/d330003/dataaccess/#} & Yes\textsuperscript{\textdagger} \\
    IFS & ECMWF ensemble tracks & \url{https://rda.ucar.edu/datasets/d330003/dataaccess/#} & Yes\textsuperscript{\textdagger} \\
    \midrule
    \multicolumn{4}{l}{\textbf{AI weather models}} \\
    \midrule
    Pangu-Weather\textsuperscript{\dag} & Global outputs & \url{https://github.com/198808xc/Pangu-Weather} & Yes \\
    FourCastNet\textsuperscript{\dag} & Global outputs & \url{https://github.com/NVlabs/FourCastNet} & No \\
    FourCastNetV2\textsuperscript{\dag} & Global outputs & \url{https://github.com/NVIDIA/torch-harmonics} & Yes \\
    AIFS & Global outputs (single member, v1.0) & \url{https://huggingface.co/ecmwf/aifs-single-1.0} & Yes \\
    GenCast & Ensemble tracks & \url{https://deepmind.google.com/science/weatherlab} & Yes\textsuperscript{\textdagger} \\
    FNV3P2 & WeatherNext 2 Cyclones ensemble tracks & \url{https://deepmind.google.com/science/weatherlab} & Yes\textsuperscript{\textdagger} \\
    \midrule
    \multicolumn{4}{l}{\textbf{NHC ATCF models (North Atlantic only)}} \\
    \midrule
    SHIP & Statistical--dynamical intensity forecasts & \url{https://ftp.nhc.noaa.gov/atcf/archive/} & No \\
    HMON & Ocean-coupled hurricane model forecasts & \url{https://ftp.nhc.noaa.gov/atcf/archive/} & No \\
    HWFI & Interpolated HWRF forecasts & \url{https://ftp.nhc.noaa.gov/atcf/archive/} & No \\
    \midrule
    \multicolumn{4}{l}{\textbf{Related datasets}} \\
    \midrule
    TC PRIMED & ML-ready TC dataset & \url{https://rammb-data.cira.colostate.edu/tcprimed/} & No \\
    SHIPS & SHIPS developmental (predictor) dataset & \url{https://rammb2.cira.colostate.edu/research/tropical-cyclones/ships/development_data/} & No \\
    \bottomrule
\end{tabular}
    \vspace{2mm}
    \captionsetup{justification=raggedright,singlelinecheck=false}
    \caption{Data Sources Summary. \textsuperscript{\textdagger}Non-gridded data only\\{\footnotesize\textsuperscript{\dag}Model run using ECMWF AI-models repository at \url{https://github.com/ecmwf-lab/ai-models}}}
    \label{tab:data_sources}
\end{table}

\begin{table}[htbp]
    \centering
    \begin{tabular}{|ll|ll|ll|}
        \hline
        SID & Basin & SID & Basin & SID & Basin \\
        \hline
        2022355S10128 & SI & 2023193N37305 & NA & 2023261N13248 & EP \\
        2023005S18142 & SP & 2023194N16123 & WP & 2023265N29284 & NA \\
        2023013S08081 & SI & 2023201N13134 & WP & 2023266N16116 & WP \\
        2023015N36289 & NA & 2023202N13240 & EP & 2023266N16323 & NA \\
        2023016S15160 & SP & 2023207N09141 & WP & 2023267N14242 & EP \\
        2023020S18156 & SP & 2023212N19090 & NI & 2023271N14144 & WP \\
        2023030N08087 & NI & 2023213N14257 & EP & 2023271N16316 & NA \\
        2023032S12091 & SI & 2023217N15256 & EP & 2023273N16073 & NI \\
        2023036S12117 & SI & 2023218N25145 & WP & 2023276N11254 & EP \\
        2023036S12163 & SP & 2023225N14248 & EP & 2023279N08157 & WP \\
        2023050S09094 & SI & 2023226N11222 & EP & 2023281N14260 & EP \\
        2023054S19117 & SI & 2023228N12259 & EP & 2023284N10330 & NA \\
        2023055S14184 & SP & 2023231N15313 & NA & 2023289N15111 & WP \\
        2023059S15149 & SP & 2023231N15327 & NA & 2023290N12256 & EP \\
        2023087S10096 & SI & 2023232N13300 & NA & 2023292N11063 & NI \\
        2023096S07135 & SP & 2023234N18128 & WP & 2023292N13309 & NA \\
        2023101N14127 & WP & 2023234N19154 & WP & 2023292S03172 & SP \\
        2023108N03160 & WP & 2023234N25271 & NA & 2023293N12089 & NI \\
        2023129N08091 & NI & 2023239N16242 & EP & 2023294N09264 & EP \\
        2023132S02088 & SI & 2023239N18144 & WP & 2023297N11277 & NA \\
        2023138N05151 & WP & 2023239N21274 & NA & 2023302N10267 & EP \\
        2023152N26274 & NA & 2023241N10158 & WP & 2023314N05149 & WP \\
        2023156N10067 & NI & 2023241N28308 & NA & 2023316S09167 & SP \\
        2023156N12135 & WP & 2023242N15339 & NA & 2023317N10094 & NI \\
        2023160N20092 & NI & 2023247N20130 & WP & 2023321N15278 & NA \\
        2023170N11323 & NA & 2023248N12257 & EP & 2023326N10240 & EP \\
        2023173N10320 & NA & 2023249N12320 & NA & 2023334N08088 & NI \\
        2023179N15256 & EP & 2023251N16334 & NA & 2023337S08165 & SP \\
        2023180N13264 & EP & 2023258N14318 & NA & 2023349N04141 & WP \\
        2023191N10262 & EP & 2023259N15225 & EP &  &  \\
        \hline
    \end{tabular}
    \caption{SID and basin for tropical cyclones in the 2023 evaluation set. North Atlantic (NA), Eastern North Pacific (EP), Western North Pacific (WP), North Indian (NI), South Indian (SI), Southern Pacific (SP)}
    \label{tab:list_cases}
\end{table}

% \color{blue}
% \arrayrulecolor{blue}
\renewcommand{\arraystretch}{1.8}
\begin{table}[ht]
    % \color{blue}
    \centering
    \footnotesize
    % \noindent\hspace{-2cm}\makebox[\textwidth][c]{
    %     \begin{minipage}{\textwidth}
    %     \centering
        \begin{tabular}{|w{c}{2.7cm}|w{c}{0.425cm}|w{c}{0.2cm}|w{c}{0.2cm}|w{c}{0.2cm}|w{c}{0.2cm}|w{c}{0.2cm}|w{c}{0.2cm}|w{c}{0.2cm}|w{c}{0.2cm}|w{c}{0.2cm}|w{c}{0.2cm}|w{c}{0.2cm}|w{c}{0.2cm}|w{c}{0.2cm}|w{c}{0.2cm}|w{c}{0.2cm}|w{c}{0.2cm}|}\hline
        \multirow{2}{*}{\makecell{Environmental\\Variable}}&\multicolumn{17}{c|}{\textbf{Pressure Level (hPa)}}\\
        \cline{2-18}
        % \cline
        &1000&975&950&925&900&850&800&700&600&500&400&300&250&200&150&100&50\textsuperscript{*}\\\hline
        Relative Vorticity &\checkmark&\checkmark&\checkmark&\checkmark&\checkmark&\checkmark&\checkmark&\checkmark&\checkmark&\checkmark&\checkmark&\checkmark&\checkmark&\checkmark&\checkmark&\checkmark&\checkmark\\\hline
        Relative Humidity &\checkmark&\checkmark&\checkmark&\checkmark&\checkmark&\checkmark&\checkmark&\checkmark&\checkmark&\checkmark&\checkmark&\checkmark&\checkmark&\checkmark&\checkmark&\checkmark&\checkmark\\\hline
        \makecell{Geopotential\\Height} &\checkmark&\checkmark&\checkmark&\checkmark&\checkmark&\checkmark&\checkmark&\checkmark&\checkmark&\checkmark&\checkmark&\checkmark&\checkmark&\checkmark&\checkmark&\checkmark&\checkmark\\\hline
        Vertical Velocity &\checkmark&\checkmark&&\checkmark&&\checkmark&&\checkmark&\checkmark&\checkmark&\checkmark&\checkmark&\checkmark&\checkmark&\checkmark&\checkmark&\\\hline
        \makecell{Horizontal\\Divergence} &\checkmark&&&\checkmark&&\checkmark&\checkmark&\checkmark&\checkmark&\checkmark&\checkmark&\checkmark&\checkmark&\checkmark&&\checkmark&\checkmark\\\hline 
        \makecell{Equivalent\\Potential\\Temperature} &\checkmark&&&\checkmark&&\checkmark&&\checkmark&\checkmark&\checkmark&\checkmark&\checkmark&\checkmark&\checkmark&&&\\\hline 
        Zonal Wind (u)&\checkmark&&&\checkmark&&\checkmark&&\checkmark&&\checkmark&&\checkmark&\checkmark&\checkmark&&&\\\hline
        Meridional Wind (v)&\checkmark&&&\checkmark&&\checkmark&&\checkmark&&\checkmark&&\checkmark&\checkmark&\checkmark&&&\\\hline
        \end{tabular}
        \renewcommand{\thetable}{S\arabic{table}}
        \caption{Pressure level environmental fields potentially useful for TC intensity prediction. Adapted from~\citep{ganesh2023}.}
        \label{tab:pl_fields}
        % \end{minipage}}
\end{table}

\begin{table}[ht]
    % \color{blue}
    \centering
    \footnotesize
    % \noindent\hspace{-2.5cm}\makebox[\textwidth][c]{
        % \begin{minipage}{1.15\textwidth}
        % \centering
        \begin{tabular}{|c|c|}
        \hline
        Temperature (2m) & Dew point temperature (2m) \\\hline 
        Convective available potential energy & Sea surface temperature \\\hline
        Total column water vapor & Total column cloud ice water \\\hline
        Total column cloud liquid water & Total column super-cooled liquid water \\\hline
        Total column cloud rain water & Vertical integral of divergence of cloud frozen water flux \\\hline
        Vertical integral of divergence of cloud liquid water flux & Vertical integral of divergence of mass flux \\\hline
        Vertical integral of divergence of moisture flux & Vertical integral of divergence of total energy flux \\\hline
        Vertical integral of potential and internal energy & \\\hline \end{tabular}
    \renewcommand{\thetable}{S\arabic{table}}
    \caption{Single level environmental fields potentially useful for TC intensity prediction. Adapted from~\citep{ganesh2023}. }
    \label{tab:sl_fields}
    % \end{minipage}}
\end{table}

\arrayrulecolor{black}

\textbf{Atmospheric Field Data.} We use atmospheric field data from a variety of sources, including:
\begin{itemize}
    \item Climate reanalysis and forecast datasets including ERA-5, TIGGE, and IFS.\\
    \item Outputs from AI models, including PanguWeather, FourcastNet, and the single member version of AIFS 1.0.
\end{itemize}
  % \begin{itemize}
  %   \item Climate reanalysis and forecast datasets including ERA-5, TIGGE, IFS, and GFS.
  %   \item Outputs from AI models, including PanguWeather, FourcastNet, and the single member version of AIFS 1.0.
  % \end{itemize}

% \textbf{Textual Data:}
%   \begin{itemize}
%     \item Research papers and reports relevant to tropical cyclone data.
%   \end{itemize}

% \textbf{TCPRIMED:}
%   \begin{itemize}
%     \item A dictionary mapping for TC PRIMED data structures to other indices.
%     \item TRACKS: Specialized time series data with entries formatted as \texttt{timestamp, intensity, pressure, latitude, longitude}.
%   \end{itemize}

% \subsection{Dataset Tree Structure}
% \begin{verbatim}
% matched_tracks/
%     2023_aifs.csv
%     ...
% neural_weather_models/
%     model_name/
%         storm_YYYYMMDD.nc
%         ...
% \end{verbatim}

% \begin{itemize}
%     \item \texttt{matched\_tracks/} contains track data aligned with observations.
%     \item \texttt{neural\_weather\_models/} contains raw AI model outputs corresponding to observed storms.
% \end{itemize}
% \begin{itemize}
%     \item Organized in a hierarchy: \texttt{SOURCE/FOLDER/DATA\_GROUPS/VARIABLE\_NAME/files}.
%     \item Naming convention for files: \texttt{ModelPrefix\_year\_varname.extension}.
%     \item Potential inclusion of CSV files within the source folder.
%   \end{itemize}

\subsection{Data Workflow}
The TCBench workflow standardizes diverse data sources into a common format for evaluation:
\begin{enumerate}
    \item Forecast and reanalysis fields are subset in space (storm-centered region) and time (forecast lead times) using storm initialization from IBTrACS.
    \item Model-specific track outputs (e.g., TIGGE XML files, neural weather model forecasts in netCDF format) are converted into a uniform CSV format that includes the storm identifier (taken from IBTrACS),  position, maximum sustained wind, and minimum sea-level pressure. See \Cref{tab:aifs2023} for an example, noting that probabilistic/ensemble models further include an ``ensemble\_idx'' column.
    \item A \texttt{TCTrack} object is created for each storm, encapsulating observed and predicted values across time steps.
    \item Gridded predictors (e.g., neural weather model forecasts) are stored as multidimensional arrays \texttt{[samples, time, lat, lon, variables]} for use in ML-based experiments.
\end{enumerate}

{

\begin{table}[ht]
    \centering
    \renewcommand{\thetable}{S\arabic{table}}
    
    \footnotesize
    \begin{tabular}{lllcccc}
        \toprule
        \textbf{sid} & \textbf{initial\_time} & \textbf{valid\_time} & \textbf{wind\_max} & \textbf{pressure\_min} & \textbf{lat} & \textbf{lon} \\
        \midrule
        2022355S10128 & 2023-01-01 00:00:00 & 2023-01-01 06:00:00 & 26.31 & 991.58 & $-$17.25 & 127.50 \\
        2022355S10128 & 2023-01-01 00:00:00 & 2023-01-01 12:00:00 & 19.08 & 993.44 & $-$17.75 & 128.00 \\
        ... & & &  & & & \\
        \bottomrule
    \end{tabular}
    \renewcommand{\thetable}{S\arabic{table}}
    \caption{Example deterministic cyclone forecast data. Probabilistic data further includes an ``ensemble\_idx'' column.}
    \label{tab:aifs2023}
\end{table}
}

\subsection{Data Processing}
Each dataset undergoes preprocessing to ensure comparability:
\begin{itemize}
    \item\textbf{TIGGE ensembles:} Extract ensemble mean and member tracks, harmonize to IBTrACS temporal resolution.\\
    \item\textbf{AI models:} Post-process global forecasts to storm-centered tracks; derive intensity and MSLP comparable to IBTrACS. \\
    \item\textbf{Post-processing Models:} Extract 50 member ensemble from parametric distribution through sampling, clip predictions to physical values ($V_{max}\geq0\;kt$).
\end{itemize}
% \begin{itemize}
%     \item \textbf{TIGGE ensembles:} Extract ensemble mean and member tracks, harmonize to IBTrACS temporal resolution.
%     %\item \textbf{ERA5:} Subset storm-centered domains (20° $\times$ 20° boxes), interpolate to common resolution, and align with best-track times.
%     \item \textbf{AI models:} Post-process global forecasts to storm-centered tracks; derive intensity and MSLP comparable to IBTrACS.
%     \item \textbf{Post-processing Models:} Extract 50 member ensemble from parametric distribution through sampling, clip predictions to physical values ($V_{max}\geq0\;kt$).
%     % \item \textbf{Extended Best Track:} Variables such as wind radii are included when available for size/intensity verification.
%     %\item \textbf{SHIPS:} Developmental data include both forecasts and environmental predictors (e.g., vertical wind shear, SST, potential intensity). Forecasts are compared against IBTrACS intensities, while the predictor set provides standardized features for ML-based benchmarking.
% \end{itemize}

\subsection{Model Evaluation and Metrics}
% We evaluate forecast performance using the following metrics, standard in tropical cyclone verification:
% \begin{itemize}
%     \item\textbf{Track error:} Great-circle distance (km) between forecast and observed positions.\\
%     \item\textbf{Intensity error:} Mean absolute error (MAE) of maximum sustained winds (kt) and minimum sea-level pressure (hPa).\\
%     \item\textbf{Skill scores:} Relative to climatology, persistence, and SHIPS baselines.\\
%     \item\textbf{Probabilistic metrics:} Brier score and reliability diagrams for ensemble/AI probability forecasts.\\
%     \item\textbf{Partitioning:} Cross-validation by year ensures temporal independence between training and evaluation sets.
% \end{itemize}
% Tom: Skipping the list of evaluation metrics given that it is given in appendix E
Standardized track files are evaluated in two steps. First, every 2023 IBTrACS record at 00 or 12~UTC defines a forecast case (storm, initial time) with valid times every 6~h up to 120~h, and the persistence and climatology baselines are initialized at the same times. Each model's track file is matched to these cases, and per-case scores and RI flags are written to separate deterministic, probabilistic, and RI result files. Second, these results are averaged per lead time and over all lead times, with ensemble member errors averaged per case and along- and cross-track errors averaged as absolute values. They are then shown, together with each model's coverage, in summary figures (all lead times) and scorecards (a subset of lead times), where RMSE and CRPS are expressed as percentage differences from persistence. $R^2$ and the area under the ROC curve for RI are also computed but not shown.

\subsection{Biases and Limitations}
\begin{itemize}
    \item The dataset represents a subset of all possible TC tracks; some regions and seasons may be underrepresented, especially where fewer storms materialize.\\
    \item Forecast evaluation may omit rare teleconnection patterns or unusual environmental scenarios, limiting generalizability.\\
    \item Definitions of storm intensity (e.g., 1-min vs. 10-min sustained winds) vary by agency and observation method, introducing potential inconsistencies across basins. We control for this by relying on USA reported values.\\
    \item IBTrACS quality varies by basin, with lower reliability in the South Indian Ocean and in the pre-satellite era.\\
    \item Reanalysis products such as ERA5 contain uncertainties in poorly observed regions and for weak or short-lived systems.\\
    \item AI models are rapidly updated (e.g., FourCastNet v1 vs. v2 vs. v3); benchmark results reflect the versions available to the authors at the time of dataset creation.\\
    \item Storm size and structural parameters are inconsistently available across sources, limiting evaluation beyond track and intensity.  
\end{itemize}
% \begin{itemize}
%     \item The dataset represents a subset of all possible TC tracks; some regions and seasons may be underrepresented, especially where fewer storms materialize.
%     \item Forecast evaluation may omit rare teleconnection patterns or unusual environmental scenarios, limiting generalizability.
%     \item Definitions of storm intensity (e.g., 1-min vs. 10-min sustained winds) vary by agency and observation method, introducing potential inconsistencies across basins. We control for this by relying on USA reported values.
%     \item IBTrACS quality varies by basin, with lower reliability in the South Indian Ocean and in the pre-satellite era.
%     \item Reanalysis products such as ERA5 contain uncertainties in poorly observed regions and for weak or short-lived systems.
%     \item AI models are rapidly updated (e.g., FourCastNet v1 vs. v2 vs. v3); benchmark results reflect the versions available to the authors at the time of dataset creation.
%     \item Storm size and structural parameters are inconsistently available across sources, limiting evaluation beyond track and intensity.
% \end{itemize}

%\subsection{Observations}
% Describe observation sources (e.g., IBTrACS, ERA5, satellite).

%\subsection{Model Forecasts}
% Detail both physics-based and ML-based forecast sources.

%\subsection{Auxiliary Data}
% Optional: describe additional indicators or reanalysis-derived fields.

%%%%%%%%%%%%%%%%%%%%%%%%%%%%%%%%%%%%%%%%
\section{Baseline Models}
TCBench provides a set of baseline models including traditional statistical approaches, numerical weather prediction (NWP) models, and simple machine learning (ML) approaches. These baselines serve as reference points for track, intensity, and rapid intensification forecasts.

\subsection{Persistence and Climatology Baselines}
We rely on persistence and climatology as so called "naïve" baselines to provide context for the performance of the evaluated models against straightforward, transparent prediction methods.
\begin{itemize}
    \item \textbf{Persistence:} Assumes that storm motion vector, intensity, and structure remain unchanged from the current state. Provides a naive forecast useful for short-lead comparisons.\\
    \item \textbf{Climatology / Mean Tendency by lead time and basin (CLIM):} Uses historical averages of storm positions and intensities within each basin and season to generate forecasts. Captures long-term trends and typical seasonal behavior.
\end{itemize}
% \begin{itemize}
%     \item \textbf{Persistence:} Assumes that storm position, intensity, and structure remain unchanged from the current state. Provides a naive forecast useful for short-lead comparisons.
%     \item \textbf{Climatology / Mean Track – Long Baseline (MT-LB):} Uses historical averages of storm positions and intensities within each basin and season to generate forecasts. Captures long-term trends and typical seasonal behavior.
% \end{itemize}

\subsection{Numerical Weather Prediction (NWP) Baselines}
We rely on physics-based Numerical Weather Prediction models as baselines for global predictions of TCs. These models rely on physics and numerical solutions of known equations to predict the state of the atmosphere based on initial conditions. We use ECMWF/TIGGE IFS in TCBench, which are ensemble forecasts from ECMWF’s Integrated Forecasting System, for both track and intensity benchmarks. We take the data provided by TIGGE as it is openly available and due to computational limitations associated with running ensembles of physics-based NWP models.

\subsection{Neural Weather Model Baselines}
We rely on neural weather models available at the time of writing to provide baselines for global weather prediction using artificial intelligence. A brief description of the models is given below:

\begin{itemize}
    \item \textbf{FourCastNetV2} - neural weather model that relies on spherical harmonics based neural operators to learn a grid-invariant evolution of the state of the atmosphere.\\
    \item \textbf{PanguWeather} - neural weather model that relies on an earth-specific transformer architecture and multiple, lead time specific models to predict the evolution of the atmosphere. \\
    \item \textbf{AIFS Single 1.0} - neural weather model whose architecture relies on a graph-based latent representation of the earth and a sliding window transformer processor in order to predict the evolution of the atmosphere. \\
    \item \textbf{Weathernext Gen (GenCast)}\textsuperscript{\textdagger} - neural weather model that relies on a conditional diffusion architecture to predict a possible evolution of the state of the atmosphere.\\
    \item \textbf{FNV3P2}\textsuperscript{\textdagger} -  Weathernext 2 Cyclones \cite{Alet2026}, relies on a functional generative network (FGN)~\citep{alet2025skillfuljointprobabilisticweather} and encodes IBTrACS information to better predict tropical cyclones.
\end{itemize}
% \begin{addmargin}[1em]{0em}
%     $\bullet\;$\textbf{FourCastNetV2} - neural weather model that relies on spherical harmonics based neural operators to learn a grid-invariant evolution of the state of the atmosphere.\vspace{\listspacing}\\
%     $\bullet\;$\textbf{PanguWeather} - neural weather model that relies on an earth-specific transformer architecture and multiple, lead time specific models to predict the evolution of the atmosphere. \vspace{\listspacing}\\
%     $\bullet\;$\textbf{AIFS Single 1.0} - neural weather model whose architecture relies on a graph-based latent representation of the earth and a sliding window transformer processor in order to predict the evolution of the atmosphere. \vspace{\listspacing}\\
%     $\bullet\;$\textbf{Weathernext Gen (GenCast)}\textsuperscript{\textdagger} - neural weather model that relies on a conditional diffusion architecture to predict a possible evolution of the state of the atmosphere.\vspace{\listspacing}\\
%     $\bullet\;$\textbf{FNv3}\textsuperscript{\textdagger} - neural weather model that is not currently open, but is purported to be a functional generative network (FGN)~\citep{alet2025skillful} fine-tuned on IBTrACS to better predict tropical cyclones.    
% \end{addmargin}
Track and intensity forecasts for the models marked with \textsuperscript{\textdagger} are taken from Google WeatherLab (\url{https://deepmind.google.com/science/weatherlab}). 
% \begin{itemize}
%     \item \textbf{NCEP GFS / GEFS:} Global Forecast System provides deterministic forecasts, while GEFS provides ensemble reforecasts from TIGGE. Baselines include position, maximum wind, and minimum sea-level pressure.
%     \item \textbf{ECMWF IFS / TIGGE IFS:} Ensemble forecasts from ECMWF’s Integrated Forecasting System. Used for both track and intensity benchmarks.
% \end{itemize}

\subsection{Postprocessing Baselines}
We provide a set of postprocessing baseline models following~\cite{gomez2025global}. Each postprocessing model uses square, clipped neural weather model fields centered on the position of the tropical cyclone at the initial time of forecast, with an extent of +/- 30\textdegree latitude and longitude. These models further use an embedding of the position, maximum wind speed, and minimum sea level pressure at the initial time of forecast for their prediction. These models are set up for distributional regression of the rate of the intensification, and output the mean and standard deviation for a gaussian distribution associated with forecast. Of particular note, the postprocessing models only predict intensification, and do not predict the motion of the storm (i.e., the evolution of the position of the TC). 

We provide three postprocessing models with varying levels of computational complexity:

\begin{itemize}
    \item \textbf{Multiple Linear Regression:} the neural weather model forecasts are reduced (e.g., by calculating extrema) and a pair of MLRs are trained to predict the mean and standard deviation of the rate of intensification.\\
    \item \textbf{Multilayer Perceptron (ANN):} the neural weather model forecasts are reduced in a manner consistent with the reduction used for the MLR, but the regression algorithm is instead a MLP.\\
    \item  \textbf{UNet:} the neural weather model forecasts and an embedding of the scalar quantities associated with the forecast (e.g., the position and intensity at the initial time of forecast) are used in the regression task.
\end{itemize}
% \begin{addmargin}[1em]{0em}
%     $\bullet\;$ \textbf{Multiple Linear Regression:} the neural weather model forecasts are reduced (e.g., by calculating extrema) and a pair of MLRs are trained to predict the mean and standard deviation of the rate of intensification.\vspace{\listspacing}\\
%     $\bullet\;$ \textbf{Multilayer Perceptron (ANN):} the neural weather model forecasts are reduced in a manner consistent with the reduction used for the MLR, but the regression algorithm is instead a MLP.\vspace{\listspacing}\\
%     $\bullet\;$  \textbf{UNet:} the neural weather model forecasts and an embedding of the scalar quantities associated with the forecast (e.g., the position and intensity at the initial time of forecast) are used in the regression task.
% \end{addmargin}
We use the details provided for these architectures and associated hyperparameters provided by~\cite{gomez2025global} in training the models.

% \begin{itemize}
%     \item \textbf{Linear Regression / Ridge Regression:} Maps current storm features (location, intensity, environmental fields) to future track and intensity.
%     \item \textbf{Random Forests:} Non-linear regression trees to capture feature interactions for both track and intensity predictions.
%     \item \textbf{Multilayer Perceptrons (MLP):} Feedforward neural networks trained on SHIPS-style predictors to forecast intensity.
%     \item \textbf{Convolutional or 3D CNNs:} When gridded environmental fields (e.g., ERA5) are available, these models predict intensity and rapid intensification events.
% \end{itemize}

% Describe baseline models: NWP, ML, persistence, climatology, etc.

\subsection{Training Setup}
All baselines are trained and evaluated consistently:
\begin{itemize}
    \item \textbf{Data Splits:} Storms are split into training, validation, and test sets by year (and not by season) to minimize temporal leakage, ensuring that models are evaluated on unseen storms.
    \item \textbf{Input Variables:} Depending on the baseline, inputs include:
        \begin{itemize}
            \item Current storm state: latitude, longitude, maximum wind, minimum pressure
            \item Environmental predictors: derived from ERA5 or TIGGE
            \item Historical storm statistics (for climatology)
        \end{itemize}
    \item \textbf{Ensemble Strategy:} For probabilistic evaluation, ML ensembles produce multiple realizations; NWP ensembles (e.g., GEFS, IFS) are used directly as provided; Post-processing model output distributions are sampled to produce a 50-member ensemble.
\end{itemize}
Baseline forecasts are evaluated using the same deterministic, probabilistic, 
and rare event metrics described in Section~4.

% \subsection{Baseline Models Summary Table}
\begin{table}[ht]
\centering
\small
\begin{tabular}{p{3cm}p{3cm}p{4cm}p{3cm}}
\toprule
\textbf{Baseline} & \textbf{Type} & \textbf{Inputs} & \textbf{Outputs / Target} \\
\midrule
Persistence & Deterministic & Current storm state & Track, intensity \\
Climatology / Mean Tendency (CLIM) & Statistical & Mean storm values, aggregated by lead time and basin & Track, intensity \\
% NCEP GFS / GEFS & NWP / Ensemble & Model initial conditions & Track, intensity \\
TIGGE IFS & NWP / Ensemble & Model initial conditions & Track, intensity \\
FourCastNetv2 & AI & Model Initial Conditions & Track, intensity\\
PanguWeather & AI & Model Initial Conditions & Track, intensity\\
AIFS Single & AI & Model Initial Conditions & Track, intensity\\
WeatherNext Gen & AI & N/A\textsuperscript{*} & Track, intensity\\
FNV3P2 & AI & N/A\textsuperscript{*} & Track, intensity\\
Pangu+MLR & PostProcessing AI & AI Model Forecasts, Current Storm state & Intensity\\
Pangu+ANN & PostProcessing AI & AI Model Forecasts, Current Storm state & Intensity\\
Pangu+UNet & PostProcessing AI & AI Model Forecasts, Current Storm state & Intensity\\

% Linear Regression / Ridge & ML & Current storm + environmental predictors & Track, intensity \\
% % Random Forests & ML & Current storm + environmental predictors & Track, intensity \\
% MLP & ML & SHIPS-style predictors & Intensity \\
% 3D CNN / ConvNet & ML & Gridded fields (ERA5) & Intensity / RI \\
\bottomrule
\end{tabular}
\renewcommand{\thetable}{S\arabic{table}}
\captionsetup{justification=raggedright,singlelinecheck=false}
\caption{Summary of baseline models included in TCBench. Each baseline serves as a reference for model evaluation. \textsuperscript{*}Forecasts taken directly from Google WeatherLab}
\label{tab:si_baselines}
\end{table}
% Include information on data splits, hyperparameters, ensemble strategies.

%%%%%%%%%%%%%%%%%%%%%%%%%%%%%%%%%%%%%%%%
\section{Evaluation Metrics \label{sec:eval_metrics}}
To test the forecast performance, TCBench uses both deterministic and probabilistic metrics. Deterministic evaluation focuses on the accuracy of predicted intensity and track positions. Standard regression metrics include the root mean square error (RMSE), mean absolute error (MAE), and the coefficient of determination ($R^2$). Track errors are quantified using direct positional error (DPE), cross-track error (CTE), and along-track error (ATE), which decompose the forecast position error along and across the observed storm motion.

For probabilistic forecasts, we evaluate ensemble predictions using the Continuous Ranked Probability Score (CRPS), which measures the discrepancy between the forecast distribution and the observed outcome, capturing both ensemble accuracy and spread. For track forecasts, CRPS is generalized by replacing absolute differences with Haversine distances between forecast and observed storm centers. Additional probabilistic metrics include the Brier Skill Score (BSS), which provides a complementary assessment of calibration and uncertainty representation. All details of metrics and computational procedures for all metrics are provided in this section.
% The description for the evaluation metrics and protocols go here

% tom - moving this to the limitations; I don't think this needs to be a dedicated section given that the class imbalance only comes in for certain types of problem framing (at a fixed spatial location, what is the probability of a storm forming? given a track, what is the probability of RI? etc.)
% \subsubsection{Class Imbalance in TC Studies}

% TODO: Discussion on the effect of problem framing on the relative imbalance in the target for evaluation. This is important when discussing, e.g., rapid intensification. 

\subsection{Deterministic Metrics}
\begin{itemize}
    % Regression Metrics
    \item Root Mean Square Error (RMSE): Measures the square root of the average squared differences between predicted and actual values, indicating the model's prediction accuracy.
    \item Mean Absolute Error (MAE): Represents the average absolute differences between predicted and actual values, reflecting the magnitude of errors in predictions.
    \item R-squared Score (R\(^2\)): Quantifies the proportion of the variance in the dependent variable that is predictable from the independent variables, indicating the model's explanatory power.
    \item {\textbf{Direct Positional Error (DPE)} is the most basic measure of the positional accuracy of a tropical cyclone (TC) forecast. It is defined as the distance between the forecast position $\mathbf{f}_{vt} = (\phi_{f,vt}, \lambda_{f,vt})$ and the observed position $\mathbf{ib}_{vt} = (\phi_{ib,vt}, \lambda_{ib,vt})$ of the storm at the same verification time.
% \begin{itemize}
    % \item 
    This metric computes the shortest distance over the Earth’s surface between the forecast and observed positions, treating Earth as a sphere. The \textit{Haversine formula} is commonly used:
    \[
    d = 2r \arcsin\left(\sqrt{\sin^2\left(\frac{\Delta\phi}{2}\right) + \cos(\phi_{ib,vt})\cos(\phi_{f,vt})\sin^2\left(\frac{\Delta\lambda}{2}\right)}\right)
    \]
    where:
    \begin{itemize}
        \item $\phi_{ib,vt}$, $\phi_{f,vt}$ are the latitudes of the observed and forecast points (in radians),
        \item $\Delta\phi = \phi_{f,vt} - \phi_{ib,vt}$, $\Delta\lambda = \lambda_{f,vt} - \lambda_{ib,vt}$ are the latitude and longitude differences,
        \item $r$ is the Earth's radius (typically $\approx$ 6371 km).
    \end{itemize}
    This method accounts for Earth's curvature and is the standard in operational TC verification.

    % \item 
    % \item \textbf{Cartesian Projection-based DPE:} An alternative approach involves projecting both observed and forecast positions onto a local tangent plane (e.g., using an azimuthal equidistant or equirectangular projection). The DPE is then computed using standard Euclidean distance:
    % \[
    % d = \sqrt{(x_{\text{fcst}} - x_{\text{obs}})^2 + (y_{\text{fcst}} - y_{\text{obs}})^2}
    % \]
    % where $(x, y)$ are the projected coordinates. This method may be preferred for regional studies or error decomposition in zonal/meridional directions.
% \end{itemize}
    \textbf{Note:} DPE provides a scalar magnitude of error but does not convey directionality (e.g., north/south or ahead/behind). Directional errors are captured by the ATE and CTE.}
    \item{\textbf{Along Track Error (ATE) and Cross Track Error (CTE)}: The DPE can be decomposed into the component along the motion of the storm---ATE---and the component perpendicular to the motion of the storm---CTE. 
    
    To calculate the ATE and CTE, we consider the position of the storm observed one 6-hourly time step prior to the verification time, $\mathbf{ib}_{ref} = (\phi_{ib,ref}, \lambda_{ib,ref})$. For calculating the components, we rely on the formula below to calculate the bearing $\beta$ between points $\mathbf{a}$ and $\mathbf{b}$ (i.e., the angle of deviation between the vector defined by the two points and the direction of true north):
    \[
      \beta(\mathbf{a}, \mathbf{b}) = \operatorname{atan2}\!\Big(
    \sin(\lambda_b - \lambda_a)\cos\phi_b,\;
    \cos\phi_a \sin\phi_b - \sin\phi_a \cos\phi_b \cos(\lambda_b - \lambda_a)
  \Big)
    \]
    We calculate the heading of the storm by calculating the bearing between the observed position $\mathbf{ib}_{vt}$ and the reference position $\mathbf{ib}_{ref}$ and reversing it:
    \[
    \theta_{\mathrm{heading}} = \beta\left( \mathbf{ib}_{vt}\;,\, \mathbf{ib}_{ref}  \right) + \pi
    \]
    The bearing of the forecast position relative to this heading can then be calculated:
    \[
    \alpha = \beta \left( \mathbf{ib}_{vt} \; , \; \mathbf{f}_{vt} \right) - \theta_{\mathrm{heading}}
    \]
    The DPE can thus be decomposed into the ATE and CTE as follows:
    \begin{align*}
        \mathrm{ATE} & = \mathrm{DPE}\;\cos{\alpha}\\
        \mathrm{CTE} & = \mathrm{DPE}\;\sin{\alpha}
    \end{align*}}
\end{itemize}
%\subsection{Tracks Error Metrics}
% e.g., track error, intensity error, landfall location/timing.
\label{appendixe:track-error-metrics}

\begin{figure}[h]
    \centering
    \includegraphics[width=0.5\textwidth]{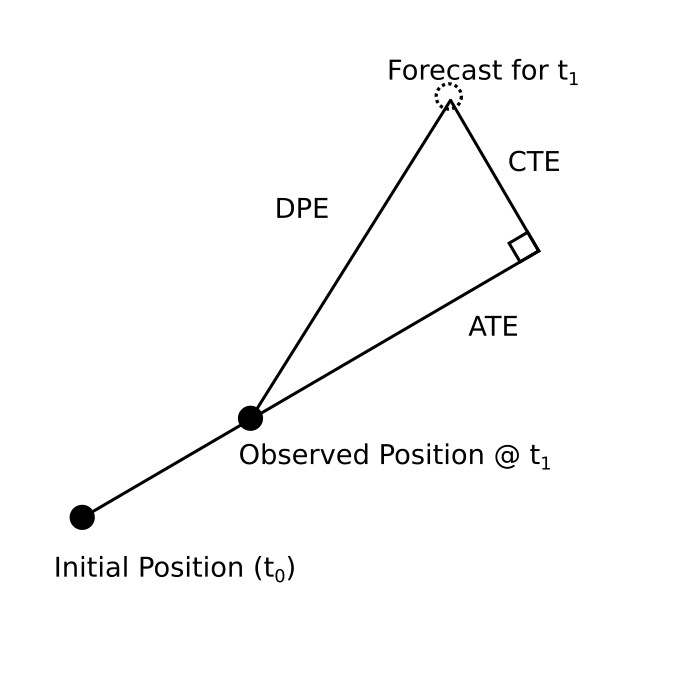}
    \setcounter{figure}{0}
    \caption{Illustration of track error metrics based on figure 5 of~\citep{heming2017tropical}. DPE: Direct Positional Error. CTE: Cross Track Error. ATE: Along Track Error.}
    \label{fig:track-error-metrics}
\end{figure}

\textbf{Cross-Track Error (CTE)} measures the component of the track forecast error \textit{perpendicular} to the observed direction of storm motion.
% \begin{enumerate}
%     \item Define the observed motion vector from the storm position 12 hours prior to VT to the observed position at VT.
%     \item Draw a perpendicular line from the forecast position to this vector.
%     \item Find the intersection point of this perpendicular with the observed vector.
%     \item CTE is the great circle distance between the forecast position and this intersection point.
% \end{enumerate}
% \begin{itemize}
%     \item In the \textbf{Northern Hemisphere}, a positive CTE indicates the forecast is to the \textbf{right} of the observed path.
%     \item In the \textbf{Southern Hemisphere}, the interpretation is reversed.
% \end{itemize}
In both hemispheres, a positive CTE indicates the forecast is to the right of the observed path.\\
% \noindent \textit{Note:} CTE is undefined at the first valid time (due to lack of a 12-hour prior observed position).

\textbf{Along-Track Error (ATE)} ATE measures the component of the error \textit{along} the direction of storm motion and reflects timing accuracy.
% \begin{itemize}
%     \item Using the same intersection point from the CTE calculation, compute the great circle distance between the observed position at VT and the intersection point.
% \end{itemize}
\begin{itemize}
    \item Positive ATE indicates the forecast is \textbf{ahead} of the observed position (i.e., storm is moving too fast).
    \item Negative ATE indicates the forecast is \textbf{behind} (i.e., storm is too slow).
\end{itemize}
We note that the use of the reference position 6 hours prior to the observation is a choice. A less noisy option is to use, e.g., using the position 12h before, with the drawback of the error being undefined near the first entry of each storm since no reference exists 12 hours prior to the first 6-hour lead time prediction. 
\subsection{Probabilistic Metrics}
% e.g., CRPS, Brier Score, Spread-Skill Ratio.
Here, we consider an ensemble of $N$ forecasts as a set denoted $\{x_i|i = 1, \ldots, N\}$, and $\{x_i\}_N$ for short.

The \textbf{Brier skill score (BSS)}~\citep{Section3_Brier_bss} is a relative measure that evaluates the improvement of a probabilistic forecast over a reference forecast or climatology. BSS ranges -\(\infty\) to 1, where 1 indicates a perfect skill and 0 indicates no improvement over the reference. A negative BSS suggests that the forecast is less accurate than the reference. The BSS is based on the Brier Score (BS), a metric used to measure the accuracy of probabilistic predictions for binary outcomes, defined as

\[
\mathrm{BS} = \frac{1}{N}\sum_{i=1}^{N}(x_i - o_i)^2
\]
where $x_i$ is a forecast probability, $o_i$ is the actual outcome, and $N$ is the number of forecasts. Then BSS is defined as
\[
\mathrm{BSS} = 1 - \frac{\mathrm{BS}_{\mathrm{forecast}}}{\mathrm{BS}_{\mathrm{reference}}}
\]

% Shortened to avoid repetition
The \textbf{Continuous Ranked Probability Score (CRPS)} is widely used to assess the skill of ensemble forecasts by measuring the discrepancy between the predicted and observed cumulative distribution functions (CDFs). Unlike the BS, which applies to binary events, the CRPS evaluates the accuracy of probabilistic forecasts for continuous variables. Formally, for a forecast CDF $\widehat{F}$ and a realized value $y$, the CRPS is
\begin{equation*}
\mathrm{CRPS} = \int_{\mathbb{R}} \left[\widehat{F}(x) - \mathcal{H}\!\left(x - y\right)\right]^2 \, dx,
\end{equation*}
where $\mathcal{H}$ denotes the Heaviside step function, which represents the degenerate CDF of a single observation. For discrete ensembles, we use the kernel representation of the CRPS~\citep{Section3_Gneiting_crps}:
\begin{equation*}
\mathrm{CRPS}\!\left(\{x_i\}_N, y\right) = \frac{1}{N}\sum_{i=1}^N |x_i - y| \;-\; \frac{1}{2N^2}\sum_{i=1}^N \sum_{j=1}^N |x_i - x_j|.
\end{equation*}
The first term is the mean absolute error of the ensemble members, while the second term measures the ensemble spread via the mean absolute pairwise difference. To mitigate finite-ensemble bias (inflated scores due to underestimation of the second term), we adopt the fair CRPS~\citep{zamo2018estimation}:
\begin{equation*}
\mathrm{fCRPS}\!\left(\{x_i\}_N, y\right) = \frac{1}{N}\sum_{i=1}^N |x_i - y| \;-\; \frac{1}{2N\left(N-1\right)} \sum_{i=1}^N \sum_{j=1}^N |x_i - x_j|.
\end{equation*}

The \textbf{probability circle} is used to evaluate the reliability of ensemble track forecasts. For each forecast, a circle is centered on the ensemble mean position $\bar{\mathbf{x}}$, computed as the normalized mean of the members' unit position vectors on the sphere. The radius $r_p$ is the $p$-quantile of the great-circle distances $d_i$ between each member and the mean of the ensemble without that member. If the ensemble spread is a reliable measure of track uncertainty, the observed position $\mathbf{y}$ should fall inside the circle with frequency $p$. The score at nominal probability $p$ is therefore the observed hit rate over $K$ cases:
\begin{equation*}
\mathrm{PC}_p = \frac{1}{K}\sum_{k=1}^{K} \mathbb{1}\!\left[\, d\!\left(\mathbf{y}_k, \bar{\mathbf{x}}_k\right) \le r_{p,k} \,\right],
\end{equation*}
where $\mathbb{1}$ is the indicator function and $d$ the great-circle distance (calculated with the haversine formula, as described in the DPE section above). Values of $\mathrm{PC}_p$ below $p$ indicate an under-dispersive (over-confident) ensemble, and values above $p$ an over-dispersive one. Following ~\cite{narita2015probability} we report $\mathrm{PC}_{70}$, the frequency with which the observed position falls within the 70\% probability circle, in \Cref{fig:si_prob_scorecard}. The values for all other deciles are calculated but not shown.

% \begin{itemize}
%     \item Brier Skill Score (BSS): Measures the improvement of the forecast probability over a reference forecast, with values ranging from -\(\infty\) to 1, where 1 indicates a perfect forecast and 0 indicates no improvement over the reference.
%     \item Continuous Ranked Probability Scores (CRPS): Quantifies the accuracy of probabilistic predictions by measuring the difference between the predicted and observed cumulative distribution functions, with lower values indicating better predictions. 
%     %\item \textcolor{blue}{Probability integral transform (PIT) histogram: Probabilistic estimate of model calibration (uniform distribution = well-calibrated; skewed or peaked distribution = poorly calibrated).--MCM}
%     %\item Interquartile range (IQR) versus error: well-calibrated models should have larger errors for larger IQRs (IQR = difference between 75th and 25th percentiles, so the middle 50\% of your predictions)--MCM}
% \end{itemize}

\subsection{Metrics for Extremes}
\textbf{Rapid Intensification:}
Rapid intensification is a rare event, but not an exceedingly rare event; it approximately corresponds to the 95th percentile of intensity change. Thus, we evaluate model performance on rapid intensification forecasts using metrics more tailored for rare event forecasts. We note that while most of TCBench is configured as a regression problem, we have formulated rapid intensification as a binary classification problem. This means rapid intensification models will be tasked with making a simple ``yes/no" prediction for the occurrence of rapid intensification. 
\begin{itemize}
 %Classification Metrics used for RI in our case
    \item Critical Success Index (CSI): Also known as the Threat Score, measures the ratio of correctly predicted positive observations to the sum of all predicted positives, actual positives, and minus true positives. %Can be used both for probabilistic track evaluation and RI. 
   % \item Mathews Correlation Coefficient (MCC): A coefficient indicating the quality of binary classifications, ranging from -1 to +1, where +1 represents a perfect prediction, 0 no better than random prediction, and -1 indicates total disagreement between prediction and observation.
\item Peirce skill score (PSS): The Peirce skill score (also known as the Hanssen and Kuipers discriminant, or the true skill statistic) is an estimate of how well the forecast separates "yes" events from "no" events. It ranges from -1 to 1, with +1 being a perfect score and 0 indicating no skill. PSS is generally considered to be better for rare events, though for extremely rare events it tends to 0.
\begin{equation}
\mathrm{PSS} \;=\; \mathrm{TPR}-\mathrm{FPR} \qquad \text{where }
\mathrm{TPR} \;=\; \frac{\mathrm{TP}}{\mathrm{TP}+\mathrm{FN}}, 
\qquad
\mathrm{FPR} \;=\; \frac{\mathrm{FP}}{\mathrm{FP}+\mathrm{TN}}.
\end{equation}
Note that in this context, the True Positive Rate (TPR) is also referred to as the Probability Of Detection (POD).
\item False Alarm Ratio (FAR): The fraction of RI forecasts that were wrong: $\mathrm{FAR} = \frac{\mathrm{FP}}{\mathrm{TP}+\mathrm{FP}}$, reported in \Cref{tab:ri_contingency}. 

%\item Odds ratio (OR): The odds ratio provides the ratio of the odds of a "yes" forecast being correct versus a "yes" forecast being incorrect. The odds ratio ranges from 0 to $\infty$, with $\infty$ being a perfect score and 1 having no skill. (We can also use the log of the odds ratio, meaning it would range from $-\infty$ to $\infty$ and 0 would indicate no skill). The odds ratio takes prior probabilities into account and is less sensitive to hedging, though it cannot be used if any section of the contingency table is 0. $ OR = \frac{hits \cdot correct negs}{misses \cdot false alarms} = \frac{\frac{POD}{1 - POD}}{\frac{POFD}{1 - POFD}}$
%\item Extreme dependency score (EDS): The extreme dependency score quantifies the association between forecast and observed rare events. Note that EDS does not tend to 0 for as event probabilities approach 0, making it useful for extreme event analysis. $EDS = \frac{2\log\left(\frac{hits + misses}{total}\right)}{\log\left(\frac{hits}{total}\right)} - 1$ 
\end{itemize}

%%%%%%%%%%%%%%%%%%%%%%%%%%%%%%%%%%%%%%%%
\section{Extended Results}

\begin{figure}[h]
  \centering
  \includegraphics[width=\textwidth]{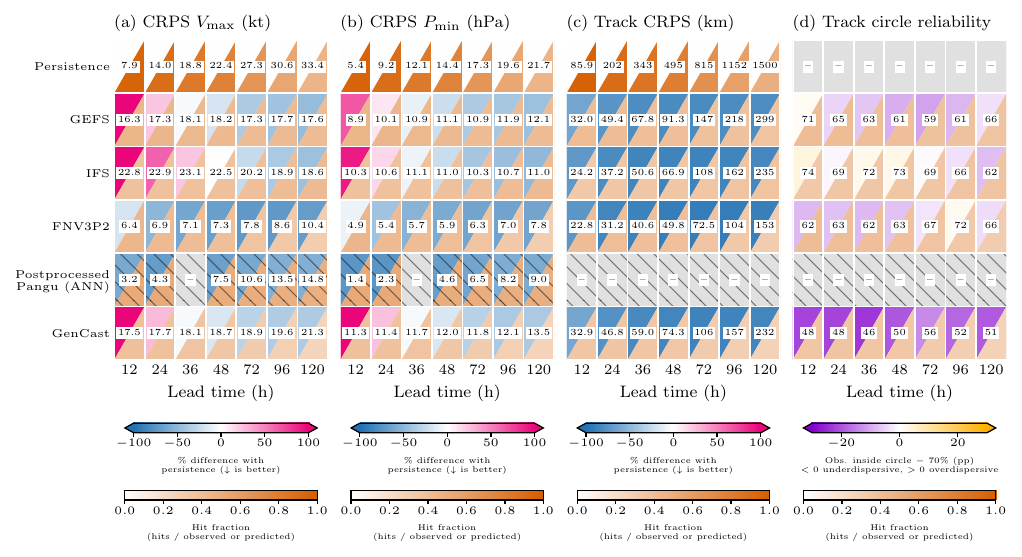}
    \caption{\textbf{Probabilistic Scorecard} - scorecards showcase a subset of predicted lead times.\\ Hatched rows indicate models that do not predict all 6-hourly lead times up to 5 days.}
  \label{fig:si_prob_scorecard}
\end{figure}

\begin{table}[htbp]
\centering
\begin{tabular}{llrrrrr}
\hline
Model & Probability & Threshold & $N$ & Base rate & TPR & FAR \\
\hline
AIFS & RI & 0.300 & 12769 & 0.111 & 0.000 & 1.000 \\
Mean Tendency (CLIM) & RI & 0.300 & 19562 & 0.072 & 0.054 & 0.172 \\
FCNet & RI & 0.300 & 6149 & 0.090 & 0.000 & -- \\
FNV3P2 & RI\_fraction & 0.300 & 8743 & 0.074 & 0.489 & 0.427 \\
GenCast & RI\_fraction & 0.300 & 7895 & 0.080 & 0.061 & 0.639 \\
Postprocessed Pangu (ANN) & RI\_fraction & 0.300 & 4386 & 0.056 & 0.687 & 0.458 \\
Postprocessed Pangu (MLR) & RI\_fraction & 0.300 & 4386 & 0.056 & 0.850 & 0.675 \\
Postprocessed Pangu (UNet) & RI\_fraction & 0.300 & 4386 & 0.056 & 0.736 & 0.723 \\
Pangu & RI & 0.300 & 6818 & 0.085 & 0.000 & -- \\
Persistence & RI & 0.300 & 19562 & 0.072 & 0.057 & 0.175 \\
GEFS & RI\_fraction & 0.300 & 8864 & 0.070 & 0.003 & 0.600 \\
IFS & RI\_fraction & 0.300 & 7998 & 0.090 & 0.001 & 0.923 \\
\hline
\end{tabular}
\caption{Rapid intensification contingency summary, all lead times pooled and considering all basins. A case is forecast as RI when the probability is at or above the threshold. Base rate: observed RI fraction of each model's cases; TPR: true positive rate; FAR: false alarm ratio. RI\_fraction refers to the number of members predicting RI over the total number of members in the ensemble.}
\label{tab:ri_contingency}
\end{table}

\subsection{Basin-Specific Evaluation - application to the North Atlantic Basin}

As an example of the utilization of TCBench to basin-specific evaluations, we provide an evaluation of the cases corresponding to the North Atlantic~(NA) basin---noting that limiting the basin to the NA requires changing a string in the configuration file and can be applied to other basins. We selected the NA basin due to the ease of retrieval of basin-specific models from the National Hurricane Center's Automated Tropical Cyclone Forecasting (ATCF) archive~\cite{NHC_ATCF_2023}. %added the reference to the model repository Dan shared
From the archive, we select SHIP, HMON, and HWFI. SHIP is a statistical–dynamical intensity-prediction scheme, HMON is an ocean-coupled dynamical hurricane model, and HWFI represents interpolated HWRF track and intensity forecasts. The evaluation also includes GEFS from the TIGGE archive; AIFS and FNV3P2 as global weather prediction systems; postprocessed Pangu (ANN) as intensity-prediction baseline; and persistence and mean tendency (CLIM) as reference baselines. The deterministic summary, probabilistic summary, deterministic scorecard, probabilistic scorecard, and RI summary for the NA basin are provided in \Cref{fig:deterministic_summary_NA}, \Cref{fig:probabilistic_summary_NA}, \Cref{fig:det_scorecard_na}, \Cref{fig:prob_scorecard_na}, and \Cref{fig:ri_summary_na}, respectively.

\Cref{fig:deterministic_summary_NA} summarises deterministic intensity and track errors by lead time, together with forecast coverage. For most models, maximum-wind and minimum-pressure MAEs increase towards 120 h. GEFS and AIFS maximum-wind MAEs are nearly flat but start from comparatively high values. Postprocessed Pangu (ANN) maximum-wind error rises steeply and exceeds persistence at the longest leads. Direct position error (DPE) also grows with lead time. Persistence and mean tendency are far worse than the dynamical and ML models (about 1700 km and 1300 km at 120 h, against roughly 750 km or less for the models). Along-track error (ATE) and cross-track error (CTE) split the position error into components parallel and perpendicular to the observed storm motion, indicating errors in translation speed (timing) and direction respectively. The coverage panel separates hits from non-comparable cases (misses, early dissipation and late dissipation). Hit counts differ widely: about 5.5k for persistence and mean tendency, between 3k and 4k for HMON, GEFS and AIFS, and only 1.7–2k for FNV3P2, Pangu (ANN) and HWFI. FNV3P2 has the largest non-comparable count (about 3.8k). Each model’s scores describe only the cases that can be evaluated, so a lower hit count does not by itself imply poorer skill. It may reflect forecast availability (e.g., fewer lead times being predicted) or track similarity, and it separates forecast accuracy from the ability to provide comparable predictions.

\begin{figure}
    \centering
    \includegraphics[width=\linewidth]{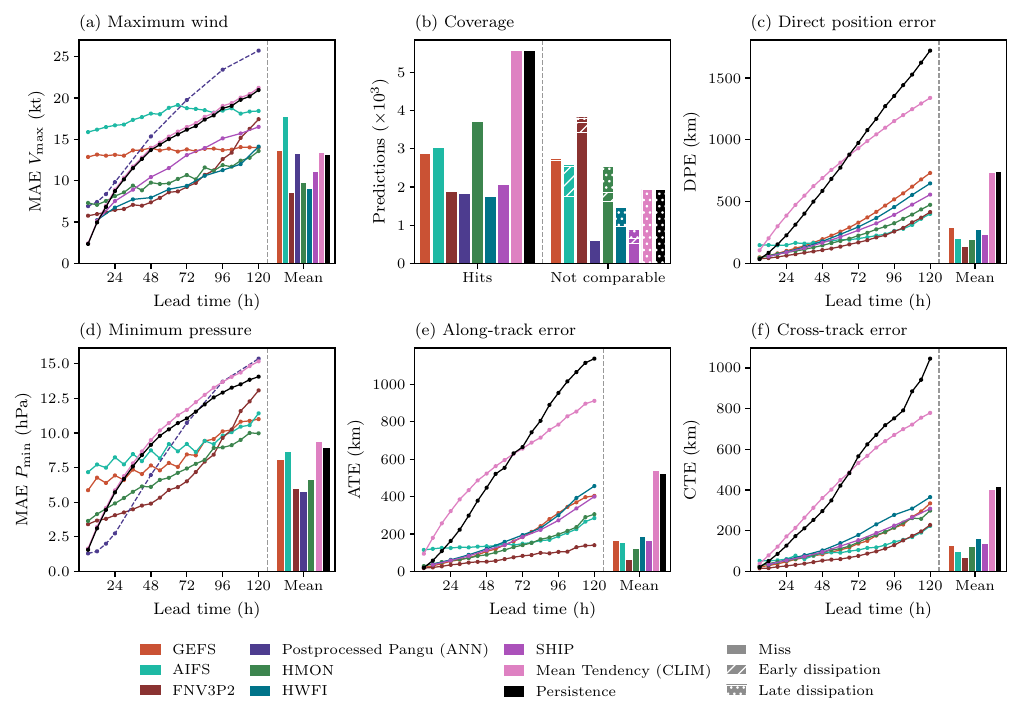}
    \caption{Summary of deterministic model performance on the 2023 set, limited to the NA basin.}
    \label{fig:deterministic_summary_NA}
\end{figure}

\Cref{fig:probabilistic_summary_NA} compares probabilistic forecasts from GEFS, FNV3P2 and postprocessed Pangu (ANN) using fair CRPS for maximum wind, minimum pressure and track, with coverage diagnostics. FNV3P2 has the lowest mean CRPS for intensity and track among the models that provide them. Its maximum-wind CRPS is below GEFS at all leads, and its track CRPS is roughly half that of GEFS throughout (about 218 km versus 405 km at 120 h). For minimum pressure, FNV3P2 has lower CRPS than GEFS through 96 h but slightly higher CRPS at 120 h. Pangu (ANN) has the lowest intensity CRPS at the earliest leads, but its CRPS grows steeply and exceeds GEFS by 96 h for maximum wind and by 72 h for minimum pressure. It provides no track CRPS and has no intensity result at 36 h. These gaps limit the scope of the comparison and are not poor scores. Because the mean-score bars are computed over different sets of available leads and cases, they should be read alongside the coverage panel.

\begin{figure}
    \centering
    \includegraphics[width=\linewidth]{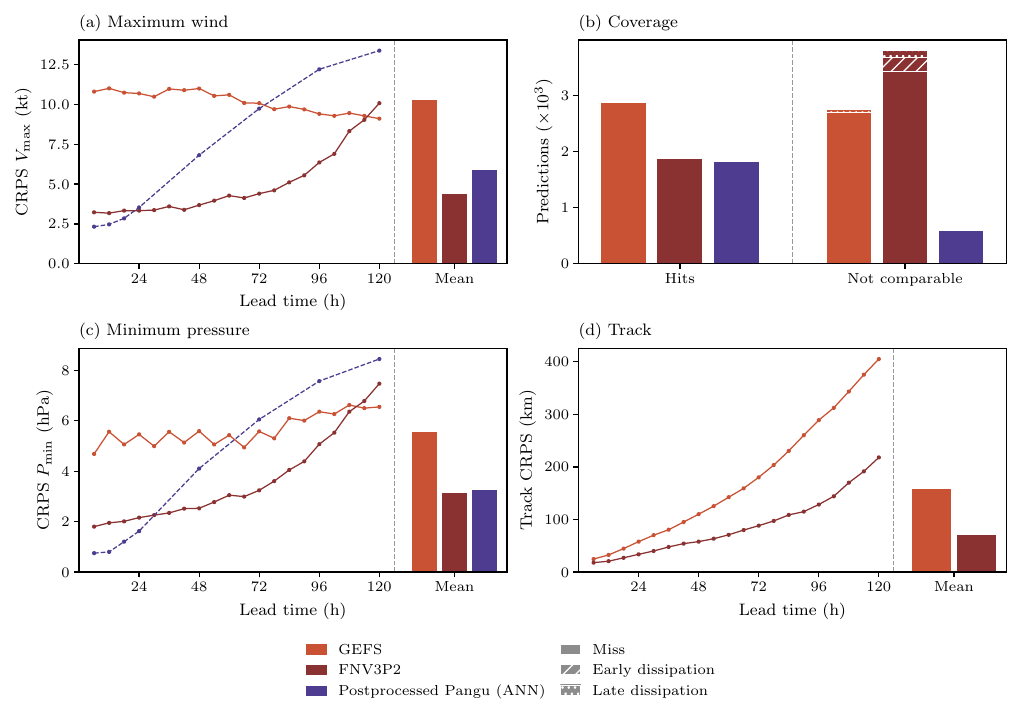}
    \caption{Summary of probabilistic model performance on the 2023 set, limited to the NA basin.}
    \label{fig:probabilistic_summary_NA}
\end{figure}

\Cref{fig:det_scorecard_na} gives RMSE for maximum wind and minimum pressure at selected lead times, together with the RI Brier skill score (BSS) relative to the mean-tendency reference. Rankings change with horizon. Among SHIP, HMON and HWFI, HWFI has the lowest maximum-wind RMSE at 24 h (9.5 kt) and HMON the lowest at 120 h (19.5 kt), which cautions against inferring a single ranking from one lead time. Minimum-pressure RMSE is unavailable for SHIP and HWFI, so missing comparisons are reported explicitly. RI BSS shows no model with consistently positive skill. Positive values are isolated: GEFS at 36–72 h and 120 h, FNV3P2 at 24 h, Pangu (ANN) at 12–24 h, and HWFI at 12 h and 48 h. HMON is negative at every lead time, and AIFS and SHIP are zero or negative.

\begin{figure}[h]
  \centering
  \includegraphics[width=\textwidth]{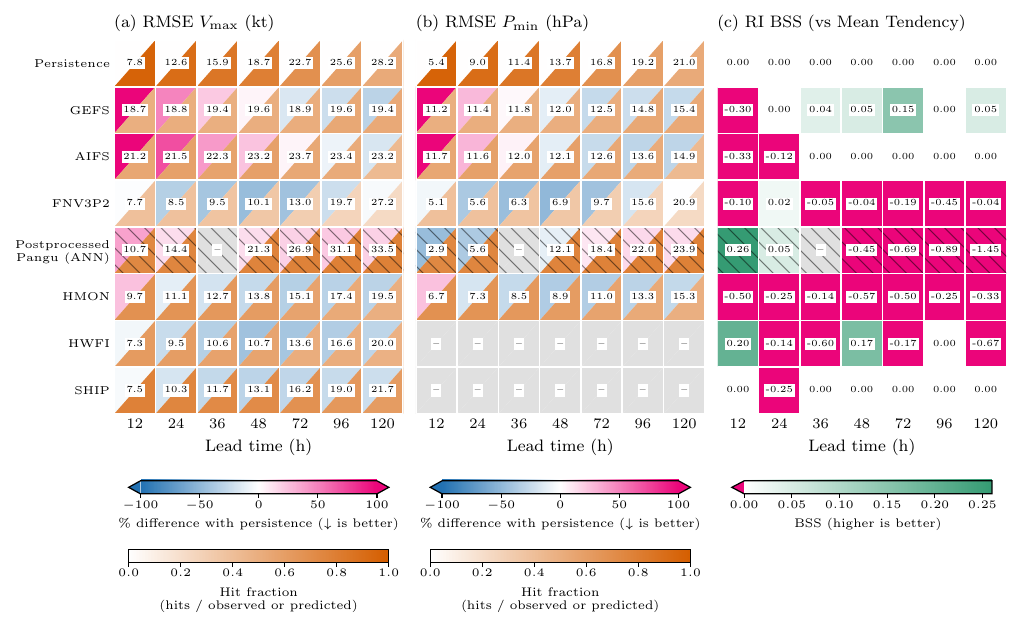}
    \caption{\textbf{Deterministic Scorecard for the NA basin} - scorecards showcase a subset of predicted lead times.\\ Hatched rows indicate models that do not predict all 6-hourly lead times up to 5 days.}
  \label{fig:det_scorecard_na}
\end{figure}

\Cref{fig:prob_scorecard_na} presents the probabilistic scorecard for the same three models as \Cref{fig:probabilistic_summary_NA}. CRPS is expressed relative to persistence, with negative percentage differences indicating lower CRPS (better). FNV3P2 beats persistence on maximum-wind and minimum-pressure CRPS at every displayed lead time. GEFS has higher maximum-wind CRPS than persistence at 12–24 h, slightly lower CRPS at 36 h, and lower CRPS from 48 h onward. For minimum pressure, GEFS is worse only at 12 h. Both models beat persistence for track CRPS, and FNV3P2 outperforms GEFS at every lead (for example, 20.7 versus 32.6 km at 12 h). The track-circle reliability panel assesses calibration through the observed fraction of cases inside the nominal 70\% circle. The observed fraction is below 70\% coverage at every displayed lead for GEFS (55–68\%) and at all leads except 96 h for FNV3P2 (71\%), with the largest shortfalls at 24–72 h. This indicates underdispersion in the track forecasts at most lead times for both models, with the degree of underdispersion varying across lead times. CRPS and circle reliability give different information: FNV3P2 achieves much lower track CRPS than GEFS without reaching its nominal coverage. The scorecard also makes missing lead times and differences in coverage visible.

\begin{figure}[h]
  \centering
  \includegraphics[width=\textwidth]{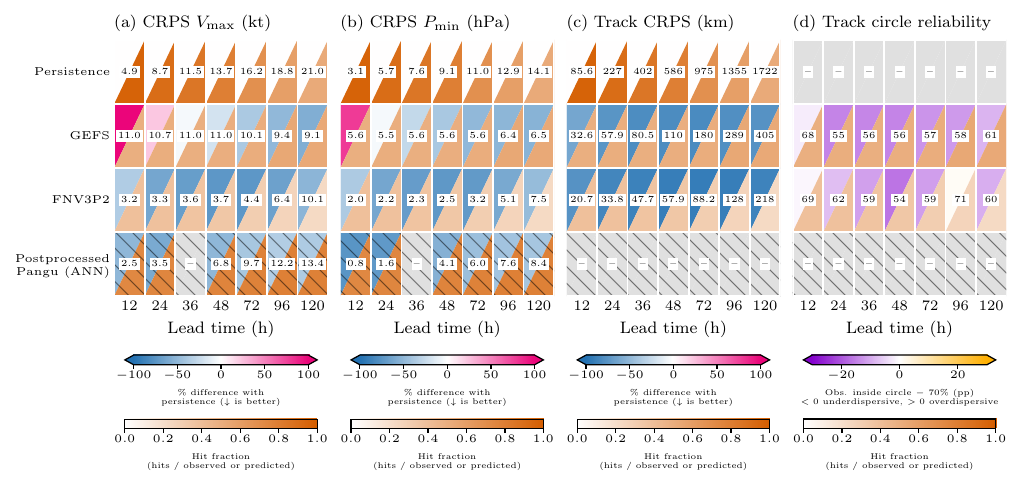}
    \caption{\textbf{Probabilistic Scorecard for NA basin} - scorecards showcase a subset of predicted lead times.\\ Hatched rows indicate models that do not predict all 6-hourly lead times up to 5 days.}
  \label{fig:prob_scorecard_na}
\end{figure}

\Cref{fig:ri_summary_na} compares RI forecasts using the critical success index (CSI) and Peirce skill score (PSS), overall and by lead time. The overall rankings differ by metric. HMON has the highest CSI (0.152), followed by HWFI (0.143) and Pangu (ANN) (0.126). Pangu (ANN) has the highest PSS (+0.405), followed by HMON (+0.249) and HWFI (+0.182). SHIP, FNV3P2, persistence and mean tendency are only marginally above zero (CSI 0.037–0.049; PSS +0.042 to +0.065), and GEFS and AIFS score near zero, meaning they show essentially no RI skill. FNV3P2 has a lower CSI than persistence (0.037 versus 0.041) but a higher PSS (+0.065 versus +0.042), another case where the metric changes the ordering. The lead-time panels show strong variation, and skill is not maintained across the five-day period. Several curves peak at the earliest leads and then drop to zero within about 24 h, and HMON fluctuates between leads. Dashed lines identify models that do not provide predictions at every 6-hourly lead time. These differences show why multiple verification metrics and lead-specific results are needed rather than a single RI ranking.

These results show how TCBench enables basin-specific evaluation across forecast accuracy, probabilistic quality, uncertainty calibration, RI skill, and forecast coverage, with rankings varying by variable, lead time, and metric.
These findings reflect the 2023 North Atlantic evaluation and may not generalize to other basins or years.

\begin{figure}[ht]
  \centering
  \includegraphics[width=\textwidth]{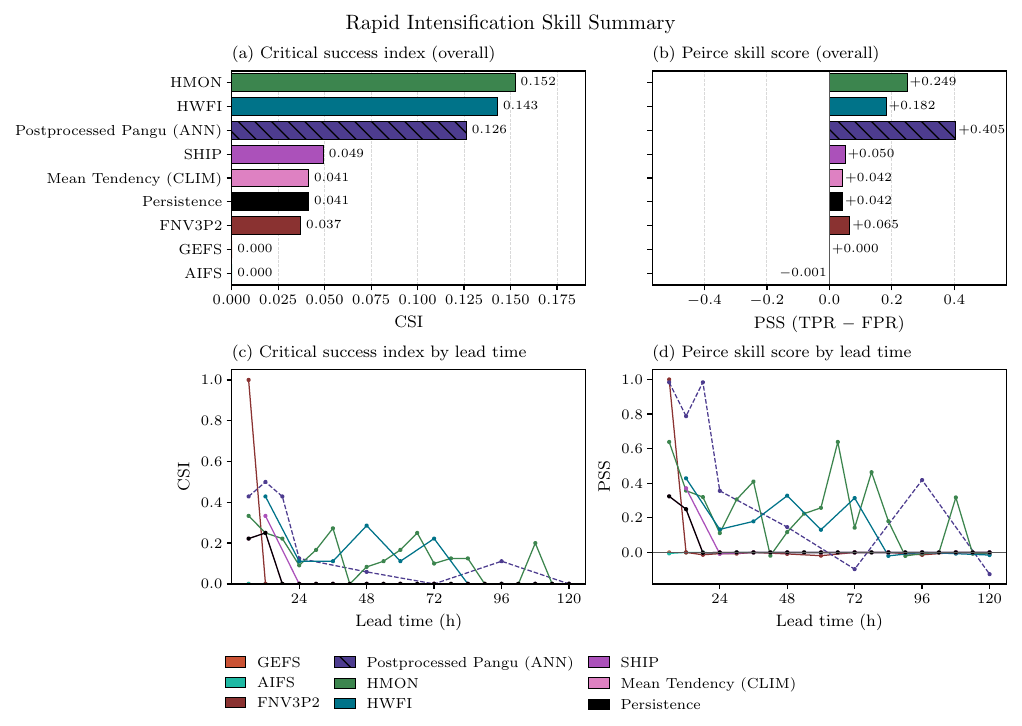}
    \caption{\textbf{RI Summary for the NA Basin} Hatched and dashed lines indicate models that do not predict all 6-hourly lead times up to 5 days.}
  \label{fig:ri_summary_na}
\end{figure}

% %========================================================
% % FIG. S3 — Rapid Intensification (overall skill by model)
% %========================================================
% \begin{figure}[h]
%   \centering
%   \includegraphics[width=\linewidth]{Evaluation_Figures/RI_skill_overall.pdf}
%   \caption{
%   \textbf{Rapid Intensification (RI) skill—overall by model.}
%   Bars show overall Critical Success Index (CSI; left) and Peirce Skill Score (PSS = TPR$-$FPR; right) 
%   computed against the IBTrACS RI ground truth for 2023. 
%   Scores use the common (SID, $t_0$, $t_0{+}L$) key set on the 6\,h IBTrACS grid.
%   }
%   \label{fig:si_ri_overall}
% \end{figure}
% \FloatBarrier
% %========================================================
% % FIG. S3 — Rapid Intensification (skill by lead)
% %========================================================
% \begin{figure}[!ht]
%   \centering
%   \includegraphics[width=\textwidth]{Evaluation_Figures/RI_skill_by_lead.pdf}
%   \caption{
%   \textbf{Rapid Intensification (RI) skill by lead time.}
%   CSI (left) and PSS (right) as a function of lead time (6–120\,h, 6\,h steps), 
%   evaluated vs. the IBTrACS 2023 RI ground truth on the 6\,h grid. 
%   Axes are capped for readability. 
%   }
%   \label{fig:si_ri_by_lead}
% \end{figure}

%%%%%%%%%%%%%%%%%%%%%%%%%%%%%%%%%%%%%%%%
% TB: Given that the deadline is Sep 24 (and not Sep 25), we cannot possibly complete this in time
% \section{Case Studies}
\FloatBarrier
\section{Potential Applications}
Accurate assessments of the wind-related risks posed by tropical cyclones rely heavily on knowledge of their full wind structures (2D images stacked in the vertical). It is challenging to obtain such wind fields because satellite observations often only provide partial coverage of TCs with limited temporal sampling. Various methods have been proposed to address this challenge, ranging from analytical parametric models~\citep{chavas2015model} to modern machine learning techniques~\citep{yang2022machine}, which attempt to infer spatial wind fields from easily observed scalar quantities such as minimum sea-level pressure or the radius of maximum surface winds. By offering IBTrACS in an AI-ready format, TCBench facilitates the design and training of models to predict key scalars needed for wind reconstruction. Moreover, the standardized aggregation of outputs from multiple AI models allows users to assess and compare tropical cyclone size predictions across models, thereby helping to identify and quantify TC size biases in neural weather models.

By providing different routines to process TC observations in different formats, the TCBench dataset is ideal for developing very short-term TC winds and precipitation predictions. Data-driven nowcasting models~\citep{agrawal2019machine} are more computationally competitive than explicitly simulating the TCs with high-resolution numerical models. The ability of probabilistic data-driven models to create hundreds of possible predictions for a given initial condition within minutes and adjust based on new satellite image inputs makes them useful additions to the existing weather forecasting pipeline. 

The examples provided in the previous sections only evaluate the intensity at a given point in the TCs. Similar to the wind field reconstruction tasks described earlier, we can develop wind downscaling models~\citep{lockwood2024generative} to add additional spatial details to the TCs in the neural weather model outputs in coarse resolution. Another extension to the prediction tasks shown in the text is to provide global prediction of TC activity across timescales. The neural weather model outputs and simple tracking algorithm in TCBench facilitate consistent evaluation of the skill of different AI models in developing the TCs at the right time and right place. Based on the comparison of TC activity patterns of different neural weather models, basin-specific refinements can be made to reduce potential spatial biases in the existing trained neural weather models. 

Finally, the aggregation of satellite images, environmental fields, and intensity observations in an easy-to-access dataset enables identification of critical predictive environmental properties for TC intensity~\citep{bister2002low}, thereby forming the basis for novel discovery on tropical cyclone physics.

\section{Limitations}
First, uncertainties associated with each product vary, which poses a challenge in estimating and providing accurate error margins for reanalyses and observations. An ambitious goal is to include observational error estimates (e.g., intensity, track location) that vary over time, reflecting improvements in satellites and technology. 

Although using multiple reanalyses and tracking algorithms is ideal due to the variation in data quality and sensitivity of cyclone detection schemes, initial efforts may need to rely on a single reanalysis dataset for tractability, while clearly acknowledging its limitations~\citep{pinheiro2020intercomparison}. Furthermore, creating a single dataset that fits all purposes is impossible, necessitating trade-offs such as neglecting some teleconnections or using datasets only available for the past approximately 15 years. Additionally, the definition of TC intensity varies by agency, primarily in the number of minutes used to calculate maximum sustained wind speeds, which introduces inconsistency.

The first position of TCs across agencies and across time is also subject to large uncertainty and inconsistency. Therefore, in subseasonal to seasonal time-scales and climate studies, it is common to consider the first time the TC reaches tropical storm intensity (35 kt) in the track to improve consistency. Moreover, creating a seamless list of predictors for the entire tropics is challenging due to the variation in relevant variables across different basins or seasons.

}

% \section*{References}

\end{document}